# Programmatic Control of a Compiler for Generating High-performance Spatial Hardware

Hongbo Rong
Parallel Computing Lab (PCL), Intel Corporation
hongbo.rong@intel.com

## Abstract

This methodology paper addresses high-performance high-productivity programming on spatial architectures. Spatial architectures are efficient for executing dataflow algorithms, yet for high-performance programming, the productivity is low and verification is painful.

We show that coding and verification are the biggest obstacle to the wide adoption of spatial architectures. We propose a new programming methodology, *T2S (Temporal to Spatial)*, to remove this obstacle. A programmer *specifies* a *temporal definition* and a *spatial mapping*. The temporal definition defines the functionality to compute, while the spatial mapping defines how to decompose the functionality and map the decomposed pieces onto a spatial architecture. The specification precisely controls a compiler to actually implement the loop and data transformations specified in the mapping. The specification is loop-nest- and matrix-oriented, and thus lends itself to the compiler for automatic, static verification. Many generic, strategic loop and data optimizations can be systematically expressed. Consequently, high performance is expected with substantially higher productivity: compared with high-performance programming in today's high-level synthesis (HLS) languages or hardware description languages (HDLs), the engineering effort on coding and verification is expected to be reduced from months to hours, a reduction of 2 or 3 orders of magnitude.

***Keywords*** High-performance computing (HPC), spatial programming, productivity, language, compiler, FPGA, CGRA

## 1 Introduction

In this paper, we address *high-performance high-productivity programming on spatial architectures.* A spatial architecture is composed of (many) distributed hardware resources, including memory blocks, arithmetic/logical elements and their interconnections. The arithmetic/logical elements execute whenever their input data are available. This assumption covers a wide spectrum of spatial architectures, from a fine-grain Field-Programmable Gate Array (FPGA) [2] to a Coarse-Grain Reconfigurable Architecture (CGRA) [3~7, 25].

In contrast with a temporal architecture (i.e. Von-Neumann architecture like CPUs or GPUs) that uses a global instruction pointer to fetch instructions from memory and then executes the instructions through a fixed pipeline, a spatial architecture has no instruction pointer or instruction fetch. Instead, instructions are directly synthesized into pipelines. This specializes the spatial architecture to match a specific computation, and thus presents better power-efficiency advantages over general purpose CPUs or GPUs.

Spatial architectures are usually used as special-purpose accelerator devices for dataflow algorithms. Dataflow algorithms are driven by data availability, which enables massive parallelism for high performance.

Performance and productivity, however, are conflicting goals. Table 1 describes several high-profile workloads that are representative of various domains and compute patterns. They include SGEMM (single-precision matrix multiply), PairHMM (Pair Hidden Markov Model), a neural network (VGG-16, mainly the convolution and ReLU layer), SpMV (Sparse-matrix dense-vector multiply) and merge sort. Figure 1 shows the engineering time spent on high-performance programming of these workloads on an FPGA or CGRA, written in several languages. The time is collected from 2 companies and 1 school based on their real-world products and research. As we can see, the productivity to achieve high performance is low. It often takes several to tens of months[1].

On temporal architectures, there is such a conflict between performance and productivity as well. But significant progress has been made to address the conflict from all perspectives of languages, compilers, libraries, runtime, auto-tuning tools and hardware [10, 11, 26, 31, 32, 51]. Particularly, the Halide language [11] well addresses the conflict in the domain of image processing, and in general, dense matrix computation.

A Halide program is a specification, including an *algorithm* and a *schedule.* The algorithm expresses a problem in a dataflow function. The schedule specifies how to optimize the function to run on hardware. Figure 2(a) illustrates the concept with a simple example, where B is a 1-dimensional (1-D) floating-point matrix, f is an arbitrary function, and i is a loop index variable iterating with unit step from 0 to some upper bound (not included). B(i) refers to the i'th element of matrix B. And f(B(i)) returns the value of function f for input B(i). The code is to compute A(i) as f(B(i)), as specified by the algorithm in Line 1~4. Line 5 is the schedule that says loop i should be fully unrolled. A compiler accepts the specification, constructs the initial loop shown in Figure 2(b) according to the algorithm, and performs unrolling according to the schedule. In general, Halide enables an expert programmer to separate the concerns of functionality and optimizations, and succinctly control a compiler to perform loop-nest optimizations. It has gained remarkable success in image processing on temporal architectures, mainly CPUs and GPUs.

Naturally, we wonder if we could adapt the philosophy of Halide, but move from temporal programming to spatial programming. A latest attempt in this direction, Halide-HLS [12], extends Halide to target FPGAs. In a graph of (different) functions, a programmer specifies a sub-graph of functions to offload to an FPGA. A programmer can optimize the communication between the functions by specifying the usage of line buffer. Each function is optimized in standard

[1] If high-performance is not the target, spatial programming does not take much time, and is not hard. Programmers can write simple loop nests and add a few pragmas to easily get average performance. This user scenario is important itself, but is beyond our scope. This paper focuses on HPC programming only.

| Workload [Source of design] | Domain | Compute patterns |
|---|---|---|
| **SGEMM** [1] | Everywhere | Massive parallel, reduction, dense matrix |
| **PairHMM** [13] | Biology | Dynamic programming, stencil |
| **VGG-16** [14] **(Convolution & ReLU)** | Machine learning | Sliding windows |
| **SpMV** [15] | Simulation, graph analytics | Sparse matrix, irregular memory access |
| **Merge sort [16]** | Database | Tree-style reduction |

**Table 1 Workloads description**

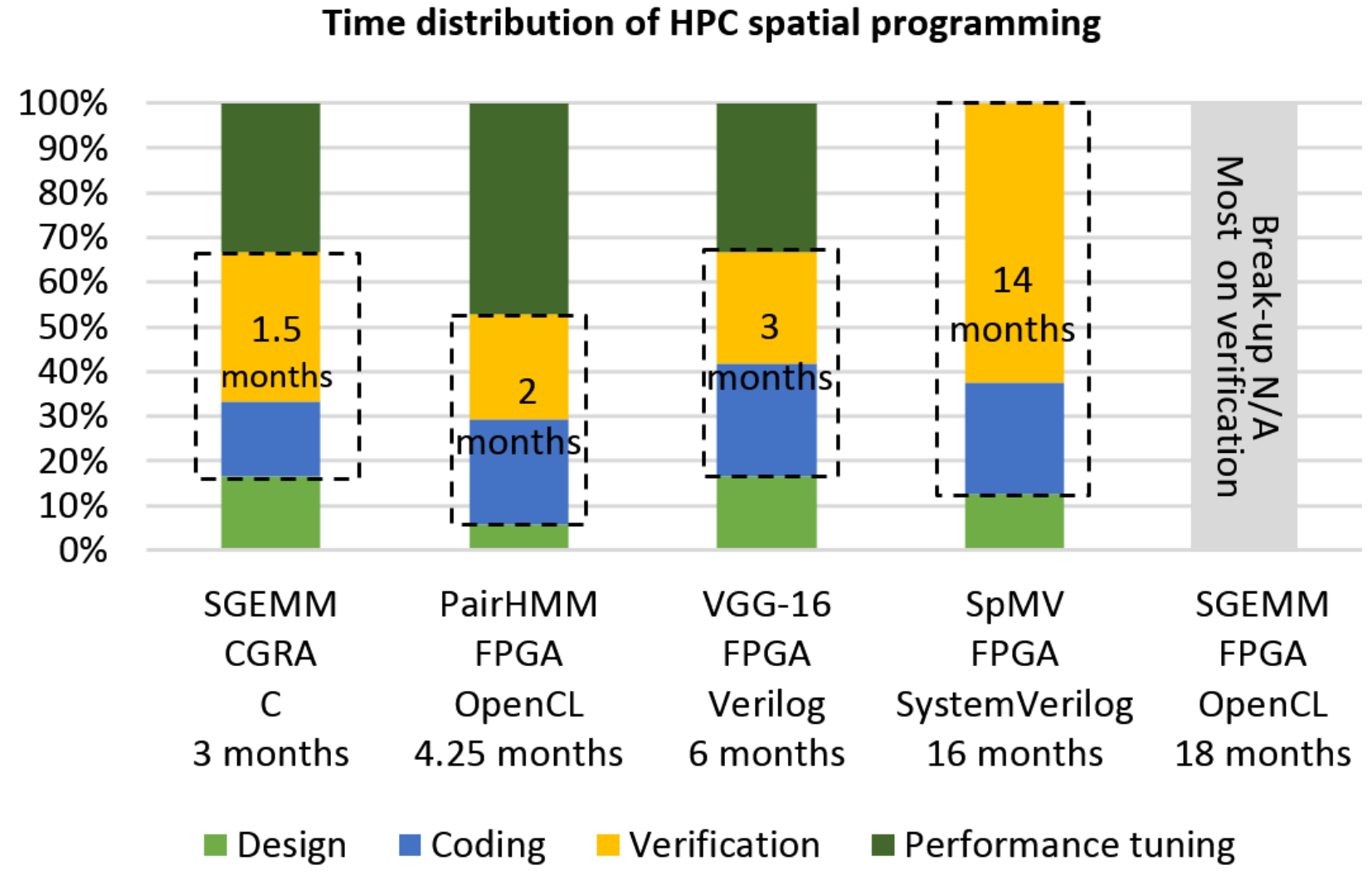

**Figure 1 HPC spatial programming in the real world[2]. Data courtesy of Daya Khudia (Intel), Gorge Powley (Intel), Yufei Ma (ASU), Jeremy Fowers (Microsoft), Davor Capalija and Tomasz Czajkowski (Intel)**

loop nest transformations and its mapping to the spatial architecture is decided by an HLS compiler automatically.

Our focus is different. We would like to exactly control the spatial mapping of a *single* function. This is a common practice in HPC programming, and is challenging, as one can see from the engineering efforts shown in Figure 1.

Usually, the workloads to accelerate on spatial architectures are simple. For example, all the workloads in Figure 1 can be simply expressed in 1 or a few mathematical equations, originally. However, a high-performance implementation can be incredibly complicated by optimizations, and have many spatial "pieces". One may get an intuition of the complexity by glancing at Figure 3(b) without understanding the details, where the workload is mathematically defined by one simple matrix expression, C = A * B, while the high-performance implementation adds 9 helper functions, and constructed 4 systolic arrays and 2 user-managed data caches. They are used to realize many loop and data transformations (Section 6.2).

Our observation is, no matter how complicated an implementation is, every spatial piece of it must be realizing a part of the functionality of the original workload, and they are communicating based on production-consumption

```
1 Parameter B(float, 1);┐
2 Var         i;        │
3 Func        A;        ├ Algorithm
4 A(i) = f(B(i));       ┘
5 A.unroll(i); ◄──────── Schedule
```

**(a) Halide code**

```
for i = 0 .. extent
    A(i) = f(B(i))
```

**(b) The loop expressed by the algorithm**

**Figure 2 A simple Halide example.** The loop bound "extent" is determined by the input matrix B's size.

relationship. So *why not first express a workload as a temporal problem, and then systematically, split it into spatial, parallel computations based on production-consumption relationship?* We call this new programming methodology *T2S (Temporal to Spatial).*

In T2S, a programmer *specifies* a *temporal definition* and a *spatial mapping*. The temporal definition defines the functionality of the original workload, while the spatial mapping defines how to decompose the functionality and map the decomposed pieces onto a spatial architecture. The specification precisely controls a compiler to actually implement the loop and data transformations specified in the mapping. This approach has the following advantages:

1. *Highly productive coding.*

   A specification is high-level and succinct. Many strategic loop and data optimizations can be specified easily. For a complex design with many loop and data optimizations, as is the common case in high-performance programming, writing a specification is far more productive than direct coding, which can be seen from the success of Halide [11] in HPC programming of CPUs and GPUs for image processing. Inspired by Halide, T2S introduces the same high productivity to spatial programming. A T2S specification, even for a complicated design, usually takes around 10s of lines (Section 6), which we expect an expert programmer to be able to write in hours instead of months.

2. *Enabling systematic expression of high-performance designs via separation of concerns.*

   A high-performance spatial design can be specified in well-defined loop and data transformations. In every specification, we separate the concerns of temporal definition and spatial mapping. As one can see from Figure 3(a) without understanding the details, a temporal definition is a dataflow function, equivalent to an un-optimized, sequential loop nest. A spatial mapping is composed of 3 parts: (1) loop transformations, which expose data locality and parallelism, (2) building a basic spatial layout, where the functionality of the transformed loop nest is split into pieces, i.e. spatial computations, connected based on their producer-consumer relationship, and (3) specializing every spatial computation, where more data and loop transformations can be applied independently.

[2] (1) The CGRA is similar to that in [7]. The FPGAs used are Stratix V for SpMV, and Arria 10 for the others.
(2) Since coding and verification phase are often intertwined, when estimation is difficult as in the case of PairHMM and VGG-16, their time is equally divided between the two.
(3) For SpMV, performance tuning did not take any significant time, as the design was carefully crafted for high-performance for the first place. The implementation was during an internship and on prototypes of the first generation of Catapult card. A lot of verification time was due to bugs or immaturity of the underlying stack, e.g. DRAM controller. An implementation on today's mature Catapult stack would be much easier.
(4) For SGEMM on FPGA, a break-up of the engineering time is not available, but most time was spent on verification. This commercial implementation took a lot of learning cost back then. Today, if without the learning cost, the total engineering time is estimated to be about 5 months.

3. *Correctness guarantee (against the temporal definition).*

A specification focuses on transforming loop nests and matrices. At such a high level, the specification lends itself to the compiler for automatic, static verification [24]. The verification ensures that the spatial mapping is semantically equivalent to the temporal definition[3]. This gets rid of the huge burden of verification from the programmer.

To understand the significance of the above advantages, let us look at Figure 1 again. As we can see, *coding and verification together consume most of programmers' time, regardless of workloads, languages and platforms.* Although our data set is small, this point likely holds in general, and is in accordance with our experiences. Verification is especially painful for software programmers: the debugging support of spatial architectures tend to be primitive, and the code is complicated by optimizations, and further complicated by timing, communication, and non-deterministic behavior. Even though performance tuning appears to be the most time-consuming phase in several case studies, it searches for the best configurations of the designs, and thus is mainly consuming machine time. It can be reduced by parallel search with more hardware resources or by auto-tuning [51]. So performance-tuning is not really a critical issue.

Therefore, by dramatically reducing the cost in coding and verification, T2S can fundamentally improve the productivity of high-performance programming on spatial architectures. We expect the coding and verification phase to be reduced from months down to hours.

T2S is generally applicable to dataflow algorithms. As long as a workload can be structured into a dataflow representation, T2S can be applied to express it (Section 4).

The paper proposes a new programming methodology. Implementing and evaluating it is the next step. We expect this paper will inspire a series of innovations in the fields of spatial programming languages and compilers.

In this paper, we assume a high-performance spatial design is provided, and our problem is how to express it. How to come up with such a design is beyond the scope of the paper, but a general principle is to fully exploit memory bandwidth and maximize computation to communication ratio [17~20].

We remark that faster evolution of designs may be enabled: with the dramatic reduction of coding and verification cost, one does not have to come up with a best design upfront, but can start from a simple design, and evolve into a high-performance one. We leave this as a future research.

Below we may refer to a "spatial architecture" or "(accelerator) device" interchangeably.

The remaining of the paper is organized as follows: Section 2 motivates the T2S approach with a simple example. Section 3 introduces the core language features and some useful extensions. Section 4 discusses the applicability of the T2S approach in terms of so-called 13 "dwarfs" (common HPC compute patterns) [22, 23]. Section 5 describes the required compiler technology. Then in Section 6, we illustrate T2S with the several workloads shown in Table 1. Section 7 compares T2S with related wok, and we reach a conclusion in Section 8.

## 2 Motivation

Let us look at Figure 2(a) again. In this Halide program, Line 1~4 specify a loop, as shown in Figure 2(b). How can we map such a temporal computation to a spatial architecture?

In Figure 4(a), we decompose the original loop quivalently into 3 pieces: They are all copies of the original loop, but specialized to implement only part of the original loop's functionality. The first piece loads B values from memory and feeds them into a 1st *channel* -- a first-in-first-out (FIFO) hardware structure on the device for fast communication; the channel is read by the middle piece for computation, which writes the resulting A values into a 2nd channel; then the 2nd channel is read out by the third piece, and the A values are stored to memory. The three pieces are running in parallel, each making forward progress whenever there are input data available.

If we name the three pieces as B_provider, A, and A_consumer, the basic spatial layout between them is shown in Figure 4(b), where every arc represents a channel.

Note that the above process preserves semantics, and is transforming a temporal computation to spatial computations.

The corresponding T2S code can be shown in Figure 4(c). Line 1~5 specify a dataflow function to compute A. It is similar to Line 1~4 of the Halide code in Figure 2(a), except we declare and set the loop bound I explicitly.

Line 6~7 form a spatial mapping by separating out the producer of B and consumer of A via two new directives "isolate_producer_chain" and "isolate_consumer_chain". The new language features will be clear in the next section.

At this moment, one can intuitively understand the code by comparing it with Figure 4(a) and (b). The code is mostly self-explanatory.

[3] As to the correctness of the temporal definition itself, it can be checked visually, or verified on CPUs functionally without worrying about the spatial architecture. This should be straightforward, as the workload expressed by the temporal definition is usually simple. So the major verification work is to verify that the spatial mapping is semantically equivalent to the temporal definition. Once that verification is passed, the compiler will generate correct spatial hardware (image) by construction.

```
1  Parameter   A(float, 2), B(float, 2), I, J, K;
2  Assumption no_alias(A, B);
3  Var          i, j, k;
4  Func         C;
5  C(i, j) = 0;
6  C(i, j) += A(i, k) * B (k, j);
7  C.set_bounds(i, 0 .. I).set_bounds(j, 0 .. J).set_bounds(k, 0 .. K);

8  Var          ii, jj, kk, iii, jjj, kkk;
9  Assumption symbolic_constants(II, JJ, KK, III, JJJ, KKK),
                divisible(I, II), divisible(II, III), divisible(J, JJ),
                divisible(JJ, JJJ), divisible(K, KK), divisible(KK, KKK);
10 C.tile(i, j, k, ii, jj, kk, iii, jjj, kkk, II, JJ, KK, III, JJJ, KKK);

11 Func A_loader, A_feeder, B_loader, B_feeder, C_collector, C_unloader,
        A_serializer(HOST), B_serializer(HOST), C_deserializer(HOST);
12  C.isolate_producer_chain(A, A_serializer, A_loader, A_feeder)
     .isolate_producer_chain(B, B_serializer,  B_loader, B_feeder)
     .isolate_consumer_chain(C, C_deserializer, C_unloader, C_collector);

13 C.unroll(ii, jj).relay(A, <0, 1>).relay(B, <1, 0>).relay(C, <-1, 0>);
14 A_serializer.remove(jjj, jj, j);
15 A_loader.remove(jjj, jj).vectorize(kkk);
16 A_feeder.unroll(ii).relay(A, <1>).buffer(A, ii, DOUBLE);
17 B_serializer.remove(iii, ii, i);
18 B_loader.remove(iii, ii).vectorize(kkk);
19 B_feeder.unroll(jj).relay(B, <1>).buffer(B, jj, DOUBLE);
20 C_collector.remove(k, kk, kkk).reorder(jj, jjj, iii).unroll(jj).relay(C, <-1>);
21 C_unloader.remove(k, kk, kkk).reorder(jj, jjj, iii);
22 C_deserializer.remove(k, kk, kkk).reorder(jj, jjj, iii);
```

Temporal definition

Spatial mapping
- Transform the loops
- Build a basic spatial layout
- Specialize individual computations

**(a) T2S code**

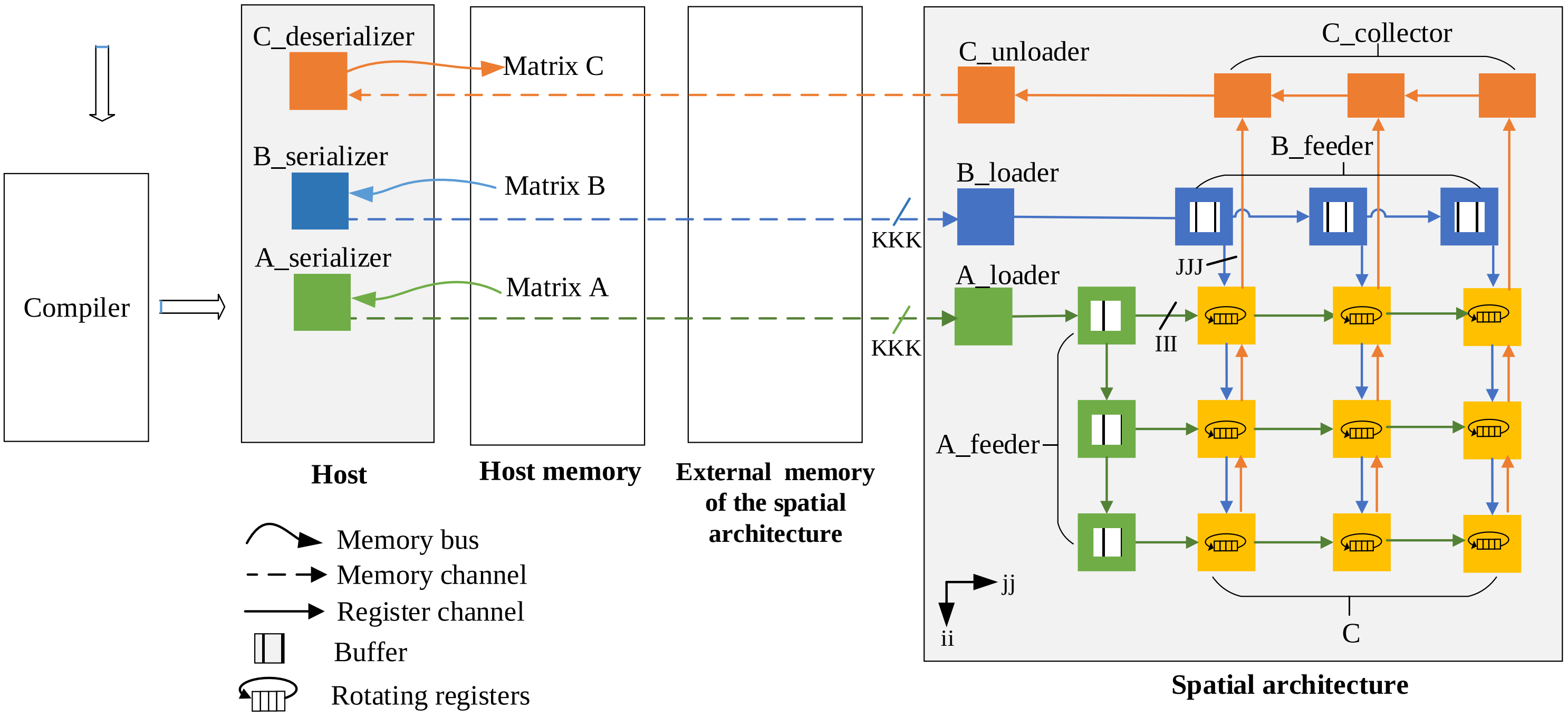

**(b) Spatial hardware and host functions**

**Figure 3. By explicitly specifying a mapping from an algorithm to a spatial architecture, a program controls a compiler to produce a sophisticated spatial hardware for SGEMM. Compared with a commercial implementation written in an HLS language for FPGA [1], the program succinctly (in ~3% of the number of lines) expresses all the strategic optimizations there. The same high performance is expected, while coding and verification of the above specification is estimated to take time in terms of hours, instead of more than 10 months.**

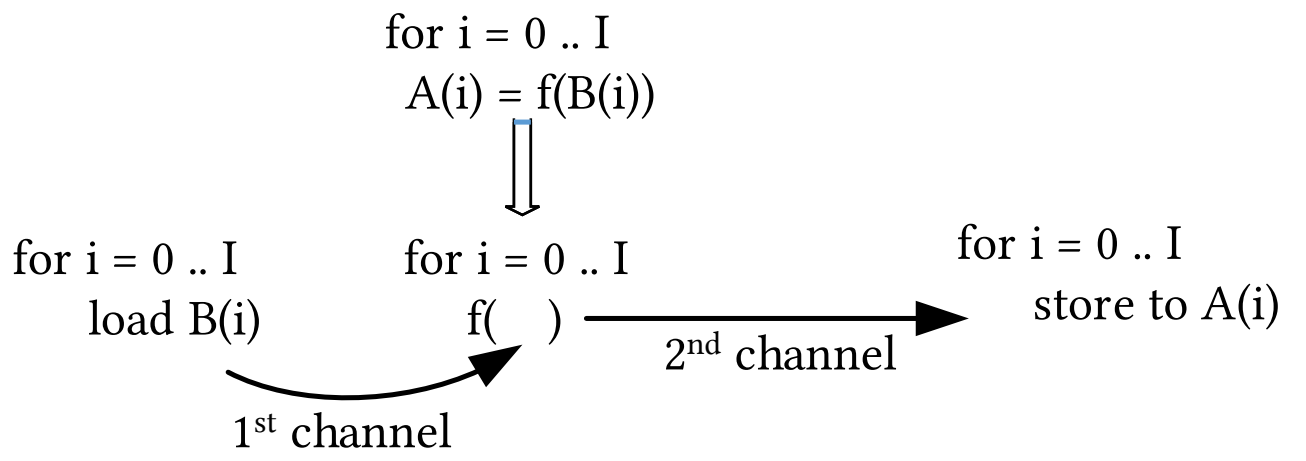

**(a) Temporal computation to spatial computations**
(corresponding to Line 1~5 of figure c below)

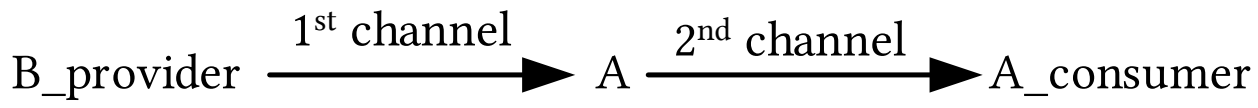

**(b) Basic spatial layout**
(corresponding to Line 6~7 of figure c below)

```
1 Parameter B(float, 1), I;          ┐
2 Var       i;                       │ Temporal
3 Func      A;                       │ definition
4 A(i) = f(B(i));                    │
5 A.set_bounds(i, 0 .. I);           ┘
6 Func      B_provider, A_consumer;  ┐ Spatial
7 A.isolate_producer_chain(B, B_provider)  │ mapping
   .isolate_consumer_chain(A, A_consumer); ┘
```

**(c) T2S code**

**Figure 4. An intuitive example.** Here I represents the upper bound of the loops.

Although not shown, the individual code pieces could be further transformed independently, as long as the semantics are still maintained. We will illustrate this point with real-world high-performance designs (Section 6).

## 3 Language features and usages

A specification is composed of *directives* that tell a compiler to represent and transform a computation. A small set of core directives can express many workloads, as listed in Table 2. Some mini-examples are embedded inside to help understanding. We inherit several directives from Halide, including Parameter, Var, Func, tile, reorder, unroll and vectorize. The other new directives we introduce are highlighted in bold fonts.

Below we introduce several important usages of the new directives. Section 6 will use real-world designs to illustrate all the usages.

1. *Static optimizations and dynamic checking*.
   Assumptions serve two purposes: static optimizations for performance and dynamic checking for correctness. The compiler may generate better-optimized code statically using the assumptions. For example, the compiler can generate cleaner tiled loops if the loop bounds are multiples of the tile sizes, as can be indicated by using the "divisible" directive.
   Using the assumptions, for correctness, the compiler may also generate host code that will check if the assumptions hold before offloading a workload to a device, to ensure meaningful results will be returned.
2. *Enabling auto-tuning*.
   A symbolic constant, as declared in an assumption, is used for performance tuning, whose value can be passed to the compiler by, for example, an auto-tuner.
3. *Constructing user-managed data cache hierarchy.*
   Creating a user-managed data cache is important for performance, especially for spatial architectures that have no hardware caches, like FPGAs. However, it is not easy to construct a cache manually. We make it easy by specifying a buffer at a given loop level. The compiler can automatically determine the buffer size and the read and write address functions (Section 5).
   For example, A.buffer(B, i) tells the compiler to build and initialize a buffer at loop level i for any B values used inside loop i. All the reads/writes of B inside loop i are redirected to reads/writes to this buffer.
   Creating buffers at more than one loop level will let the compiler to build a multi-level data cache hierarchy.
4. *Creating systolic arrays*[4].
   Unrolling, followed by data forwarding, tells the compiler to construct a systolic array. Unrolling creates a set of hardware processing elements (PEs): For a loop nest under consideration, unrolling n ($n \geq 1$) loops of it, whose number of iterations are $N_0$, $N_1$, ..., $N_{n-1}$, respectively, results in $N_0 * N_1 * ... * N_{n-1}$ number of PEs, each with a unique identifier (vector). Data forwarding connects the PEs by letting a PE sending data to another.
   With these explicit directives, the compiler implementation becomes easier: unlike a traditional compiler, a T2S compiler does not have to figure out an optimal partition [21], or how to localize dependences [29], as they have been replaced here with explicit unrolling and data forwarding, respectively.
5. *Avoiding redundant memory accesses.*
   It is important to minimize memory accesses, and thus increase the computation to communication ratio, for high performance. Loop removal removes a loop from a loop nest that contains memory references. The removed loop's index variable must not be used in the memory references, and thus the removal of the loop avoids redundant memory accesses. However, it breaks the semantics of the original loop nest. Programmers need minor changes in the consumer side of the accessed values so that the semantics keeps unchanged.
6. *Building a basic spatial layout.*
   Isolating a producer (or consumer) chain is to build a basic spatial layout. Usually, only a single producer (or consumer) is needed. However, if the produced (or consumed) data need to be preprocessed (or post-processed), an additional producer (or consumer) can be isolated, as illustrated in Row 11~12 of the table.

[4] In a systolic array, all PEs run synchronously [27]. In a wavefront array, all PEs run asynchronously and fire whenever data are available [28]. Whether a compiler generates a systolic array or wavefront array is target-dependent. To be simple, in this paper, we use "systolic array" to refer to both systolic and wavefront array.

A producer and a consumer communicate via a channel. We distinguish two kinds of channels: *memory channel* and *register channel*. A memory channel is an illusion created by the compiler automatically between the host and device, as if they communicate through a FIFO[5].

A register channel is on the device, and is composed of registers. The number of registers is called the *depth* of the channel.

7. *Serialization and de-serialization.*

   For high performance, randomly accessing the external memory of a device for input or output data is not desirable.

   For input data, a general strategy is to serialize the data on a host CPU, transfer the data to the device through a memory channel, and let the device access the data sequentially. Further, a user-managed cache can be created to store the data in the internal memory of the device, and from there, the data can be randomly accessed.

   For output data, the strategy is similar, except that the data flow from the device to the host sequentially, and it is the host to de-serialize the data into the host memory.

Beside the core directives, it can be beneficial to add some extensions for efficiency. For example, Table 3 lists some useful extension so that one may control what resulting data to store to memory, how deep a register channel is, and the usage of a line buffer.

## 4 Applicability

To understand the applicability of T2S, we looked at common compute patterns in HPC, the so-called "13 dwarfs" [22, 23]. We can classify them into 3 classes:

1. *Compute patterns that are naturally amenable to dataflow representations.*

   These patterns include dense linear algebra, dynamic programming, structured mesh, N-body, Monte-Carlo, spectral methods, and circuits. These patterns contain loops that can be easily tiled and unrolled into hardware PEs [1, 13, 14, 40~42]. These PEs are either independent or connected locally.

   We will exemplify the application of T2S to dynamic programming and structured mesh with Smith-Waterman (Section 6.1), and the application to dense linear algebra with SGEMM (Section 6.2) and convolution and ReLU (Section 6.3.1).

2. *Compute patterns that may be made as dataflow.*

   These patterns include sparse linear algebra, graph algorithms, graph models, unstructured mesh, and backtrack-branch-bound. These patterns are not natural match of dataflow representations, as they contain indirect memory accesses, data-dependent parallelism, and behavior that can only be described imperatively.

   However, for each of these patterns, we can find a high-performance case study using an overall dataflow structure on FPGAs (We use FPGAs as a proxy of spatial architectures) [15, 39, 43, 44]. For example, for SpMV, a high-performance design [15] preprocesses a sparse matrix on the host; with the preprocessed matrix as input, the workload on the device side becomes much like a regular dense matrix computation.

   So with careful designing, these compute patterns may be mapped to dataflow structures.

   Often, there are PEs in a structure that have to be programmed imperatively, for example, PEs that have internal states. As long as such PEs can be encapsulated in a dataflow interface, and their internal states are not exposed outside, the PEs can be safely used in the dataflow structure[6].

   We will exemplify the application of T2S to sparse linear algebra with SpMV (Section 6.3.2), and the application to graph algorithms with merge sort (Section 6.3.3).

3. *Finite state machine.*

   This is an imperative pattern, and not suitable for dataflow, unless it is encapsulated entirely as a PE.

Therefore, for any compute pattern to run on a spatial architecture, the real question one should ask is how to map it to a dataflow structure. As long as a dataflow structure is defined, T2S is generally applicable.

## 5 Compiler

As we said, a T2S program has a temporal definition and a spatial mapping. The compiler scans the T2S program sequentially, and thus reads the directives and implements them one by one. The temporal definition appears first. Accordingly, the compiler builds up an initial IR, i.e. an abstract syntax tree (AST) representing an un-optimized loop nest.

Then the spatial mapping appears, and is processed by the compiler as follows:

1. *Transforming the loops in the IR.*

   According to the loop transformation directives, the compiler transforms the IR. This results in an optimized loop nest that will be used as the basis for every spatial computation to create next.

2. *Building a basic spatial layout.*

   With the directives for isolating producers or consumers, the compiler splits the IR into multiple connected computations based on the producer-consumer relationship.

   As we see from Row 11~12 of Table 2, a connection is always a FIFO, i.e. a register or memory channel. A register channel is on-chip and is a widely used mechanism in HLS [46]. The compiler can automatically determine its depth. A memory channel can be constructed by sequentially writing to and reading from a virtually shared memory between the host and device. This virtually shared memory is also constructed by the compiler.

3. *Specializing the individual computations.*

[5] One should not confuse the "memory channel" in our context with a "hardware memory channel" in computer architectures, which is a hardware accessing memory.

[6] One may define such a PE as a library function, and call it in the dataflow structure. Halide supports this mechanism.

<table>
<tr><th></th><th colspan="2">Row number & Notation</th><th>Semantics</th></tr>
<tr><td rowspan="5">Temporal definition</td><td>1</td><td>Parameter p[(type, dimension)]</td><td>A runtime parameter whose value is unknown until the device runs. The parameter is a matrix if a dimension is given, otherwise a scalar. The default type is int.</td></tr>
<tr><td>2</td><td>Assumption no_alias(A, B),<br>symbolic_constant(N),<br>divisible(I, II),<br>Boolean expression</td><td>Matrix A and B do not share any memory location.<br>N is a symbolic constant.<br>I is divisible by II.<br>A Boolean expression that must be respected.</td></tr>
<tr><td>3</td><td>Var i[(type)]</td><td>Variable i, with a default type of int, if not given.</td></tr>
<tr><td>4</td><td>Func A, B(HOST)</td><td>Function A on the device, and function B on the host.</td></tr>
<tr><td>5</td><td>A(i, j)=...<br>A.set_bounds(i, 0 .. I)<br>.set_bounds(j, 0 .. J)</td><td>Internally, the compiler builds a loop nest[7]:<br>for i = 0 .. I<br>for j = 0 .. J<br>A(i, j) = ...<br>"i = 0 .. I" means i ∈ [0, I). Every loop's step is 1 by default.</td></tr>
<tr><td rowspan="5">Loop transformation</td><td>6</td><td>A.tile(i, ii, II)<br>A.tile(i, j, ii, jj, II, JJ)<br>A.tile(i, j, ii, jj, iii, jjj, II, JJ, III, JJJ)<br>...</td><td>Tile loop i with size of II into a new loop i and ii, then tile loop ii with size of III into a new loop ii and iii, etc. The order of the loops before and after tiling is as specified in the parameters. For example,<br>for i = 0 .. I<br>for j = 0 .. J<br>A(i, j) = ...<br>⟹ A.tile(i, j, ii, jj, II, JJ) ⟹<br>for i = 0 .. I step II<br>for j = 0 .. J step JJ<br>for ii = 0 .. min(I - i, II)<br>for jj = 0 .. min(J - j, JJ)<br>A(i + ii, j+ jj) = ...</td></tr>
<tr><td>7</td><td>A.reorder(i, j, ...)</td><td>Reorder loop i, j, ..., which must be adjacent but not necessarily in this order, so that their new order is in this order, starting from the innermost level.</td></tr>
<tr><td>8</td><td>A.unroll(i, ...)</td><td>Fully unroll loop i, .... For example,<br>for i = 0 .. 2<br>for j = 0 .. J<br>for k = 0 .. 6 step 5<br>A(i, j, k) = ...<br>⟹ A.unroll(i, k) ⟹<br>4 PEs, with each PE having an identifier (vector) of <0, 0>, <0, 5>, <1, 0>, or <1, 5>. Each PE's code is:<br>i, k = the identifier's first, second element<br>for j = 0 .. J<br>A(i, j, k) = ...</td></tr>
<tr><td>9</td><td>A.remove(i, ...)</td><td>Remove loop i, ... Feasible only when variable i, ... are not used inside.<br>for i = 0 .. I<br>for j = 0 .. J<br>A(i) = ...<br>⟹ A.remove(j) ⟹<br>for i = 0 .. I<br>A(i) = ...</td></tr>
<tr><td>10</td><td>A.vectorize(i)</td><td>Vectorize loop i so that its iterations run in parallel and in lock step.</td></tr>
<tr><td rowspan="2">Compute partition</td><td>11</td><td>A.isolate_producer_chain(B, Func$_1$, Func$_2$, ..., Func$_n$)</td><td>Isolate the production of B values to Func$_1$, which sends the values through Func$_2$, ..., Func$_n$. and finally to function A, via channels. Take Figure 3(a) for example, the spatial layout before and after isolating a producer chain of A in Line 12 is<br>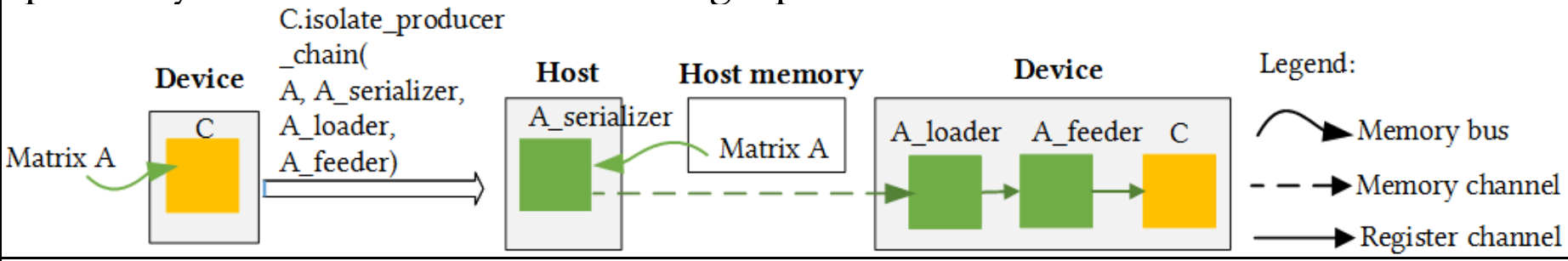
</td></tr>
<tr><td>12</td><td>A.isolate_cosumer_chain(B, Func$_1$, Func$_2$, ..., Func$_n$)</td><td>Isolate the consumption of B values to Func$_1$, which receives the values from Func$_2$, ..., Func$_n$, and finally from function A, via channels. Take Figure 3(a) for example, the spatial layout before and after isolating a consumer chain of C in Line 12 is<br>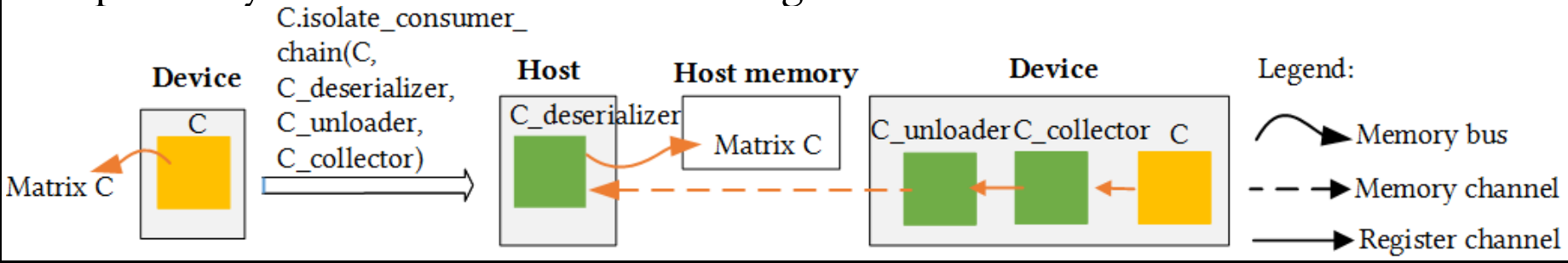
</td></tr>
<tr><td>Data caching</td><td>13</td><td>A.buffer(B[, i]<br>[, DOUBLE|REGISTER])</td><td>Create an on-chip buffer for B, at loop level i if specified and otherwise before all the loops of function A. The buffer can be double-buffer and/or implemented in registers, or by default a single buffer implemented with on-chip memory blocks.</td></tr>
<tr><td>Data forwarding</td><td>14</td><td>A.relay(B, <d1, d2, ...>)</td><td>In the systolic array of A, among the incoming B values, every PE keeps the values that belong to itself, and forwards the values that belong to other PEs to the neighbor PE whose identifier vector equals this PE's identifier vector + <d1, d2, ...>. At the boundary of the systolic array, the compiler may automatically wrap around a relay, or drop an unnecessary relay as an optimization [21].</td></tr>
</table>

**Table 2. Core directives.** We use "[]" to indicate that the content inside is optional.

[7] The IR shown here is slightly different from Halide: We assume row-major instead of column-major storage format for functions and matrices for easier understanding. Using which order is an implementation choice, and does not constrain the methodology.

| | | |
|---|---|---|
| **Temporal def.** | A.**store**(X, Y) | A(X, Y) need be stored to memory. Without it, all A values will be stored. |
| **Compute partition** | A.**set_min_depth**(B, Funcx, N) | For the register channel that sends B values from A to Funcx, set its minimum depth as N. Without this, the compiler will set a conservatively big depth. |
| **Data caching** | A.**linebuffer**(B[, i]) | Similar to buffer(), except this is a line buffer, implemented always in registers. |

**Table 3. Extension directives.**

For each spatial computation, there may be more directives given, including data caching, data forwarding, and further loop transformation, etc. The compiler correspondingly specializes an individual computation's IR.

When unrolling is specified, the compiler creates multiple copies of the IR, assigns each a unique identifier and specializes the unrolled loops' index variables with the identifier, as illustrated in row 8 of Table 2. If there was an input (or output) channel to the IR before unrolling, the channel is split into multiple channels as well, all connecting to the same source (or destination) as before unrolling. If subsequently, the source (or destination) is also unrolled, these channels will be further split. So with unrolling, one may create very sophisticated spatial hardware, like that shown in Figure 3(b). The compiler keeps track of the identifiers to ensure the right producer-consumer relationship.

In addition, the compiler automatically performs the following tasks:

4. *Static verification.*
   The compiler statically verifies that the initial and final IR are semantically equivalent. The loop-nest- and matrix-oriented specification lends itself to the compiler for automatic, static verification [24].
5. *Low-level and target-specific optimizations.*
   Besides the high-level optimizations directed by the T2S code, the compiler can transparently performs traditional low-level and target-specific optimizations.
   For example, the compiler may software pipeline a loop, minimize the depth of a channel, remove redundant or dead code, etc. The compiler may "infinitize" a loop: e.g. in Figure 4(a), since the middle piece no long uses the loop variable i and the piece works only when its input channel has values available, the compiler may transparently replace the loop "for i = 0 .. I" in the middle piece as an infinite loop, which uses less spatial resources. For another example, the compiler may generate efficient code for reduction, which is a very common pattern: For a reduction, the compiler can generate code to use a register to accumulate the result. Only when the reduction is completely done, the compiler lets the result in the register to be sent to another place, like a channel. If the reduction is inside a loop, multiple registers may be needed and for efficiency, the compiler can organize these registers as rotating registers (i.e. registers in a cyclic structure) [36]. This will be illustrated with SGEMM in Section 6.2.
6. *Code generation.*
   The compiler generates a spatial hardware (image), and the code for host-side PEs and host-device communication.

Overall, the compiler techniques for the above tasks exist today. A possible flow for implementing a compiler is shown in Figure 5, leveraging existing techniques like Halide, LLVM and HLS or HDL compiler.

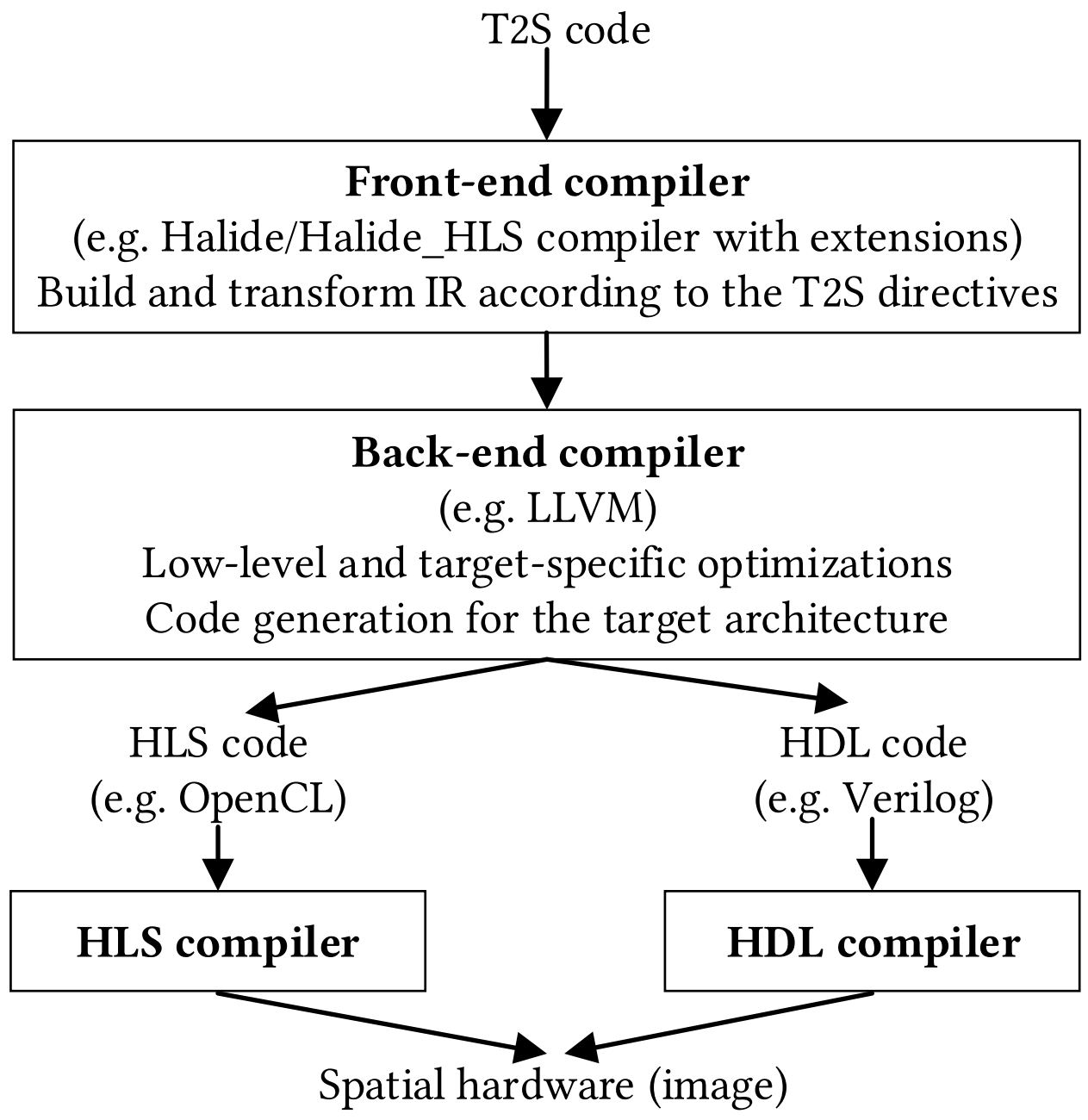

**Figure 5. A possible compiler implementation flow**

## 6 Case studies

In this section, we study in depth several representative, high-profile workloads as described in Table 1 (except we use Smith-Waterman instead of PairHMM for simplicity without losing any key points). Given a high-performance design for each of them, we show how to express the design in T2S on spatial architectures, all in ~20 lines of code or less. However, we should point out that there can be many designs for each workload, and our methodology is generally applicable, not limited to these specific designs.

These designs are for high-performance, and thus in general, have many optimizations. That is the nature of HPC programming, though. Fortunately, the optimizations they use are typical optimizations taught in college, and once understood, can be useful for many other designs. We will start simple, and guide the reader through the optimizations as gently as possibly.

### 6.1 Smith-Waterman

Smith-Waterman is a stencil computation, where to compute a data point, 3 neighbor data points are needed. The algorithm has important usage in bio-sequence alignment. It has a dependence structure similar to that of PairHMM [13], but its computation is simpler. So we use it to illustrate T2S for easier understanding.

A high-performance design for Smith-Waterman, adapted from a commercial implementation of PairHMM [13], is expressed in T2S in Figure 6(a).

First, we specify a temporal definition (Line 1~8), which tells the compiler to build an IR for a loop nest as shown in Figure 6(b), where f is a function (not shown), and S and T are bio-sequences to match. The code is self-explanatory. Note that only the last value of A is useful, indicated by "store(I - 1, J - 1)" in Line 8. So the compiler may avoid saving other A values.

Then we specify a spatial mapping as follows:

1. Transform the loops.
   Line 11 tells the compiler to tile the loop nest as illustrated in Figure 6(c). The inner two loops compare a pair of fixed-sized sub-sequences, and the outer two loops enumerate all such pairs. The inner two loops will be unrolled next. Such tiling followed by unrolling is a general strategy to decompose a big problem into sub-problems such that each sub-problem can be solved with the available hardware resources on a device.
2. Build a basic spatial layout.
   Line 13 tells the compiler to isolate a driver function as the producer of S and T, and as both the producer and the consumer of A. This results in a basic spatial layout as shown in Figure 6(d). At this moment, no function has ever been unrolled yet.
3. Specialize individual computations.
   Line 14 specializes function A. First, the innermost two loops are fully unrolled. This creates the orange-colored PEs shown in Figure 6(e), where each PE has a unique identifier annotated on top of it, corresponding to the unrolled ii and jj loop's values. The two channels originally connected with function A are automatically split by the compiler so as to connect with these PEs of function A, instead.
   Second, the A values are forwarded in 3 directions: vertical, horizontal, and diagonal. The S and T values are forwarded horizontally and vertically, respectively. In effect, the unrolled PEs with the data forwarding between them consist of a systolic array.
   What about the data forwarding at the boundary of the systolic array? Along the 3 directions, if a value is coming from outside of the systolic array, it is from the producer of it, viz. the driver, via the bold black-colored channel in Figure 6(e). Similarly, if a value is sent outside of the systolic array, it is to the consumer of it, viz. the driver as well, via a blue-colored channel in Figure 6(e).
   With existing techniques [21], the compiler can automatically connect the PEs to the right channels. Based on a producer PE's loop nest, the compiler knows the data to be sent over a channel, and thus determines the depth of the channel. For example, the compiler can figure out that at most MAX_J number of values will be accumulated in each of the blue-colored channels, and sets up their depths accordingly. Here MAX_J is a symbolic constant, an upper bound of J (Line 9).
   Finally, Line 14 also tells the compiler to create a register buffer at loop level j for A values, which encloses the two innermost loops that have become the systolic array. That is, the compiler will create a buffer big enough to provide all the input A values for one execution of the systolic array. All the input A values are redirected to this buffer, and from there to the PEs appropriately.

## 6.2 SGEMM

The problem is to compute a matrix C given matrix A and B: C= A * B. The T2S code for a commercial design [1] is shown in Figure 3(a). First, we specify a temporal definition (Line 1~7). The corresponding IR built by the compiler is shown in Figure 7(a).

Then we specify a spatial mapping as follows:

1. Transform the loops.
   Line 10 of Figure 3(a) tiles each loop two times. The IR is manipulated by the compiler correspondingly, and the result is shown in Figure 7(b).
   To understand the purpose of the tiling, look at Figure 7(c). The basic idea is, through tiling of the loops, to block the matrices; then multiply a row of A blocks with a column of B blocks to compute one C block, and repeat the same process to compute all C blocks.
   In more detail, the matrices are blocked at two levels. At the first level, the matrices are blocked in rows and columns, leading to "blocks". At the second level, each block is further blocked in rows and columns, leading to "sub-blocks". That is why the loops are tiled twice.
   The multiplication of a row of A blocks with a column of B blocks is done via rank-1 updates. That is, take the 1st column of A sub-blocks and the 1st row of B sub-blocks, which are highlighted by the green and blue backgrounds in Figure 7(c), respectively, compute their outer product, and accumulate the result into the current C block. Then take the 2nd column of A sub-blocks and the 2nd row of B sub-blocks, etc.
2. Build a basic spatial layout.
   Line 12 in Figure 3(a) separate out the functionality of the loop nest in Figure 7(b) into multiple functions, according to their producer-consumer relationship. This results in a basic spatial layout. There are 3 communication paths between the host and the device, composed of memory and register channels. We have illustrated two of the paths in Row 11~12 of Table 2. The other path for B values is similar to that for A values.
3. Specialize individual computations.
   So far, every function is a single PE itself. We can specialize the functions individually. Some of them may be specialized into systolic arrays, and some of them may get their loops transformed further.

a) Specialize function C.
   First, unroll the loop nest of function C along two middle loops ii and jj (Line 13 of Figure 3(a)). This creates a 2-dimensional systolic array, the orange PEs in Figure 3(b). All the PEs, however, are accessing memory, which is not efficient. Instead, a PE at the left boundary of the systolic array can load A values from memory, and forward them to its right neighbor, i.e. along the direction vector of <0, 1>. Similarly, a PE at the top boundary of the systolic array can load B values from memory, and forward them to its neighbor below, i.e. along the direction vector of <1, 0>. Also similarly, a PE at the bottom boundary of the array can forward a resulting C data to its neighbor above, i.e. along the direction vector of <-1, 0>. Such data forwarding between PEs are dictated by the "relay" directives (Line 13 of Figure 3(a)). Consequently, all the PEs in the 2-D systolic array get connected via channels.

b) Specialize helper functions for A.
These helpers consist of A_serializer, A_loader and A_feeder. They are responsible to smoothly feed matrix A's data into the 2-D systolic array. Since A_serializer is on the host side, while A_loader is on the device side, the compiler automatically creates a memory channel between them, as highlighted in a green-colored dotted line in Figure 3(b).
The memory channel is implemented by using the host and device memory: the compiler instruments A_serializer so that the stream of A values loaded by A_serializer are written into the host memory sequentially, then copied to the device memory (by the host-device communication code also generated by the compiler), which is sequentially read by A_loader. This in effect "serializes" matrix A.
There is a subtle point here: A_serializer should avoid writing the same data twice for performance. That is what Line 14 of Figure 3(a) does: all the loops not related with matrix A are removed.
Similarly, A_loader removed all such loops except loop j (Line 15 of Figure 3(a)). Loop j is kept because the device has limited on-chip memory to hold the data. So the same set of data have to be reloaded for each different loop j iteration.
Now it is clear that although a memory channel is conceptually a FIFO, the data stream from the producer may be "reloaded" in the consumer side, because it is implemented in memory.
Further, the reading of the memory channel by A_loader is vectorized at the innermost loop level (Line 15 of Figure 3(a)): because matrix A is serialized, one load can load multiple data from the contiguous memory locations.
Finally, in Line 16 of Figure 3(a), loop ii of A_feeder is unrolled into another set of PEs. Incoming data are forwarded from a PE to the next if the data do not belong to the current PE, and otherwise are kept in a double buffer in the current PE. The compiler automatically figures out to which location in the double buffer a data item should be written. This in effect "de-serializes" matrix A.

c) Specialize helper functions for B.
These functions are specialized in the same way, as specified in Line 17~19 of Figure 3(a).

d) Specialize helper functions for C.
These helpers consist of C_collector, C_unloader and C_deserializer (Line 20~22 of Figure 3(a)). They are responsible for smoothly draining matrix C's data into the host memory. They are also similar to A's helpers, except the data flow from the device to the host. A subtle point here is the "remove(k, kk, kkk).reorder(jj, jjj, iii)" directives in them. These directives remove the loops unrelated with C and move loop jj into the innermost level. Remember that jj is the horizontal dimension of the orange-colored 2-D systolic array (See Figure 3(b)). As such, one data item is read from one column of the 2-D systolic array, and then another data item from the next column, etc. In this way, no column of the 2-D systolic array will be stalled for too long a time in draining its data.

## 6.3 Convolution and ReLU, SpMV and merge sort

We briefly introduce how T2S is applied to these workloads, highlighting only the key points. We leave a detailed description to the Appendix.

### 6.3.1 Convolution and ReLU

Convolution layer is the most compute-intensive part in a convolutional neural network, often followed by a ReLU layer. Here we express a high-performance design in the literature [14]. Following the T2S coding template, we first specify a ReLU function, as well as a convolution function, whose output is used by the ReLU function. That is, the convolution and ReLU function have producer-consumer relationship. We have two loop nests: one loop nest for each of the two functions. First, we tile some loops of the two loop nests. Then the two loop nests (functions) are specialized individually.

For the ReLU function, two loops in it are unrolled to form a 2-D systolic array, and a buffer is added at certain loop level for caching its output.

For the convolution function, three loops in it are unrolled to form a 3-D systolic array. Two buffers, one a regular buffer and the other a line buffer, are added at an outer and an inner loop, respectively, to build a two-level data cache hierarchy, from which the PEs of the 3-D systolic array read their input.

Before unrolling, the two functions are connected via a channel. After unrolling, the channel is naturally split into many channels, connecting the 2-D and 3-D systolic array's PEs as producers and consumers. It can be complicated for human, but is mechanical for the compiler (Section 5).

### 6.3.2 SpMV

SpMV is to compute y = A * x, where x and y are dense column vectors and A is a sparse matrix. Computing patterns using sparse matrices have irregular memory access, which implies low utilization of the bandwidth of the external memory of the device. Also, the patterns have data-dependent parallelism, which means that some loops of the patterns might have dynamic trip counts. For example, in SpMV, when a row of the matrix A is multiplied with the vector x, the length of the row (i.e. the number of non-zeros in the row) is data dependent, which is not statically known.

To address these issues, a high-performance design of SpMV [15] preprocesses a sparse matrix on the host. Conceptually, after preprocessing, SpMV becomes much like a dense matrix computation, as shown below:

```
for c = 0 .. C
  for r = 0 .. NUM_SLOTS
    y(decoder(row_lengths)) += A'(c, r) * x(column_ids(c, r));
```

Here NUM_SLOTS is a compile-time constant, representing the number of physical memory slots that can be accessed in parallel from the device, C is a data-dependent runtime constant, and A' is the preprocessed matrix: multiple rows of the original matrix A have been statically scheduled into a single row of A'; there are totally NUM_SLOTS of rows in the new matrix A' – one row for a memory slot, and all the rows of A' have been padded so that they all have C number of columns. The lengths of the rows of the original matrix A have been recorded into another matrix named row_lengths. There is a new function (decoder) and another matrix (column_ids) tell from which row and column of the original matrix A, respectively, the current A' element comes from. For that purpose, the decoder function reads matrix row_lengths and deduces the current row's index.

After the preprocessing, the memory access patterns of A', column_ids, and row_lengths are regular and contiguous.

As we can see, the new loop nest has an overall dataflow structure. However, the "decoder" is not purely functional: it contains internal states. Besides, we will isolate the loading of all the matrices (A', column_ids, and row_lenghts) into a spatial piece called "matrix_fetcher", which is not exactly a dataflow actor: the matrix_fetcher fetches a data item from each of the matrices every time, and thus may finish fetching matrix row_lengths earlier as it is shorter than the other two matrices. In other words, one of the 3 inputs may no longer contain tokens at some time, but the matrix_fetcher still needs to fire (for the other two inputs). We can write the decoder and matrix_fetcher imperatively and encapsulate them as if they were dataflow functions.

We leave the details of SpMV to the Appendix.

#### 6.3.3 Merge sort

Merge sort is a reduction in tree style. Here we express the merge sorter tree in FPGASort [16]. A number of data streams are read simultaneously, and through a binary tree, merged into a single stream. At each level of the tree, the input streams stay in FIFOs (i.e. channels). The closer the level is to the root of the tree, the deeper the channels. Each tree node merges two incoming streams. Such a node has internal states and must be written in an imperative language, but it can be properly encapsulated into a dataflow function.

In T2S, all the tree nodes can be expressed as the PEs after unrolling some loops. The tree shape can be constructed by forwarding data from a node at some position m of a tree level to a node at position m/2 of the next higher tree level. This in effect constructs a tree-like systolic array. The depths of the channels at each tree level can be specified using the "set_min_depth" directive in Table 3.

## 7 Related work

Related work comes from languages and their compilers, and programming paradigms. The relevant languages include HDLs, HLS languages, and domain-specific languages (DSLs). Dataflow programming and communicating sequential processes (CSP) are relevant programming paradigms.

HDLs mainly include Verilog and VHDL. They describe a circuit at the register-transfer level (RTL) with explicit timing. They can be compared to "assembly languages".

HLS languages have a higher abstraction. An HLS program gives an algorithmic description of a desired behavior. The behavior is usually decoupled from clock-level timing. There are many HLS languages [8, 45~50]. The most common ones are based on standard languages such as C/C++/SystemC/Matlab. Source code is analyzed, architecture-constrained, and lowered down to an HDL.

Compared with a HLS program, T2S code is more succinct and at a higher abstraction level. T2S *constructively specifies* the behavior of a spatial hardware, and lets the compiler to generate details according to the specification, instead of letting the programmer directly write the details.

Another major advantage of T2S over HLS is that a T2S specification lends itself to static, automatic verification.

DSLs also have a higher abstraction level than HLS languages. Such languages express and optimize an algorithm in predefined domain-specific patterns, and lower the patterns into an HDL [11, 12, 30, 33~35]. The Halide [11] and Halide-HLS [12] DSLs have been introduced in Section 1. T2S may be implemented based on them.

Related compiler techniques have been discussed in Section 5.

Dataflow programming [38] models a computation as a direct graph, where nodes are operations, edges are channels, and data flow from nodes to nodes along the edges.

CSP applies dataflow to describe a concurrent system with component processes working independently and communicating with message passing via channels [37].

T2S can be viewed as a succinct way to construct a special CSP system, where every process (i.e. PE) is the result of isolation and specialization from a single loop nest.

## 8 Conclusion and future work

"It is remarkable how complex a simple computation can be when performance is at stake" [31]. High-performance programming is complicated by optimizations. We have shown a programming methodology, namely T2S, which substantially reduce the complexity of high-performance spatial programming by separating the concerns of temporal definition and spatial mapping. T2S enables programmers to succinctly specify strategic loop and data optimizations, and leaves to a compiler the implementation of these optimizations, the verification of them, and all the other low-level and target-specific optimizations. In this way, we show the promise of fundamentally improving the productivity of high-performance spatial programming.

We are planning for implementing a language and compiler for T2S based on Halide/Halide-HLS.

A future research is to specify a workload in a simple or familiar language (e.g. in un-optimized C/Python or Alpha [29] code), and have a tool (based on e.g. roofline or polyhedral model like PolyMage [32]) to make the decisions on loop and data transformation, and thus automatically generate T2S code – for even higher productivity.

```
1  Parameter    T(int, 1), S(int, 1), I, J;
2  Func         A;
3  Var          i, j;
4  A(-1, -1) = 0;
5  A(i, -1)  = 0;
6  A(-1, j)  = 0;
7  A(i, j)   = f(A(i -1, j), A(i, j - 1), A(i - 1, j - 1), T(i - 1), S(j - 1));
8  A.set_bounds(i, 0 .. I).set_bounds(j, 0 .. J).store(I - 1, J - 1);

9  Assumption symbolic_constant(II, JJ, MAX_J),
              divisble(I, II), divisible(J, JJ), J ≤ MAX_J;
10 Var          ii, jj;
11 A.tile(i, j, ii, jj, II, JJ);

12 Func         driver;
13 A.isolate_producer_chain(S, driver)
     .isolate_producer_chain(T, driver)
     .isolate_producer_chain(A, driver)
     .isolate_consumer_chain(A, driver);

14 A.unroll(ii, jj).relay(A, <1, 0>).relay(A, <0, 1>)
     .relay(A, <1, 1>).relay(T, <1, 0>)
     .relay(S, <0, 1>).buffer(A, j, REGISTER);
```

Temporal definition

Transform loops

Build a basic spatial layout

Specialize A

Spatial mapping

**(a) T2S code**

```
for i = 0 .. I
 for j = 0 .. J
   if (i == 0 && j == 0) A(-1, -1) = 0
   if (j == 0) A(i, -1) = 0
   if (i == 0) A(-1, j) = 0
   A(i, j) = f(A(i - 1, j), A(i, j - 1), A(i - 1, j - 1), T(i − 1), S(j − 1))
```

**(b) Sequential loop nest** (corresponding to Line 1~8 in figure a )

```
for i = 0 .. I step II     } Tiling
 for j = 0 .. J step JJ    }
  for ii = 0 .. II
   for jj = 0 .. JJ
    i', j' = i + ii, j + jj
    if (i' == 0 && j' == 0) A(-1, -1) = 0
    if (j' == 0) A(i', -1) = 0
    if (i' == 0) A(-1, j') = 0
    A(i', j')=f(A(i'-1, j'), A(i', j'-1), A(i'-1, j'- 1), T(i' − 1), S(j' − 1))
```

**(c) Transformed loop nest** (after Line 11 in figure a)

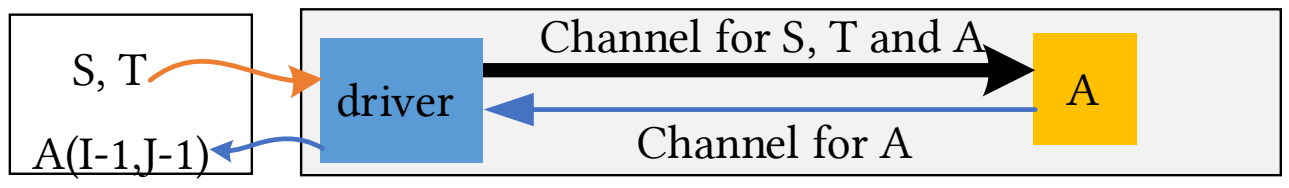

**(d) Basic spatial layout** (after Line 13 in figure a)

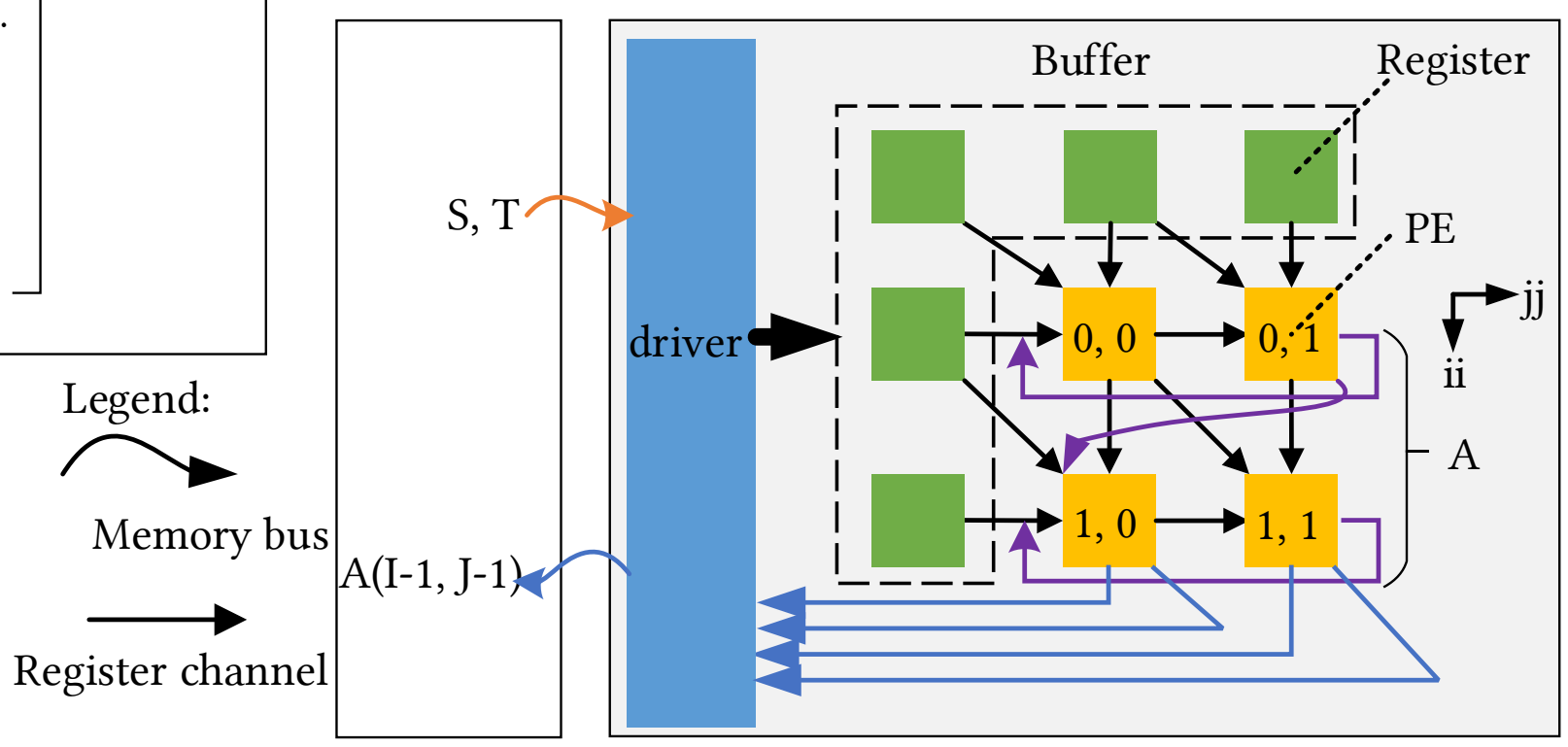

**(e) Spatial hardware** (after Line 14 in figure a)

**Figure 6 Smith-Waterman**

```
for i in 0 .. I
 for j in 0 .. J
  for k in 0 .. K
   if (k == 0) C(i, j) = 0
   C(i, j) += A(i, k) * B(k, j)
```

**(a) The sequential loop nest**
(corresponding to Line 1~7 in Figure 3(a))

```
for i in 0 ..  I step II         }
 for j in 0 .. J step JJ         } First level tiling
  for k in 0 .. K step KK        }
   for ii in 0 .. II step III    }
    for jj in 0 .. JJ step JJJ   } Second level tiling
     for kk in 0 .. KK step KKK  }
      for iii in 0 .. III
       for jjj in 0 .. JJJ
        for kkk in 0 .. KKK
         i', j', k'= i+ii+iii, j+jj+jjj, k+kk+kkk
         if (k' == 0) C(i', j') = 0
         C(i', j') += A(i', k') * B(k', j')
```

**(b) Transformed loop nest** (after Line 10 in Figure 3(a))

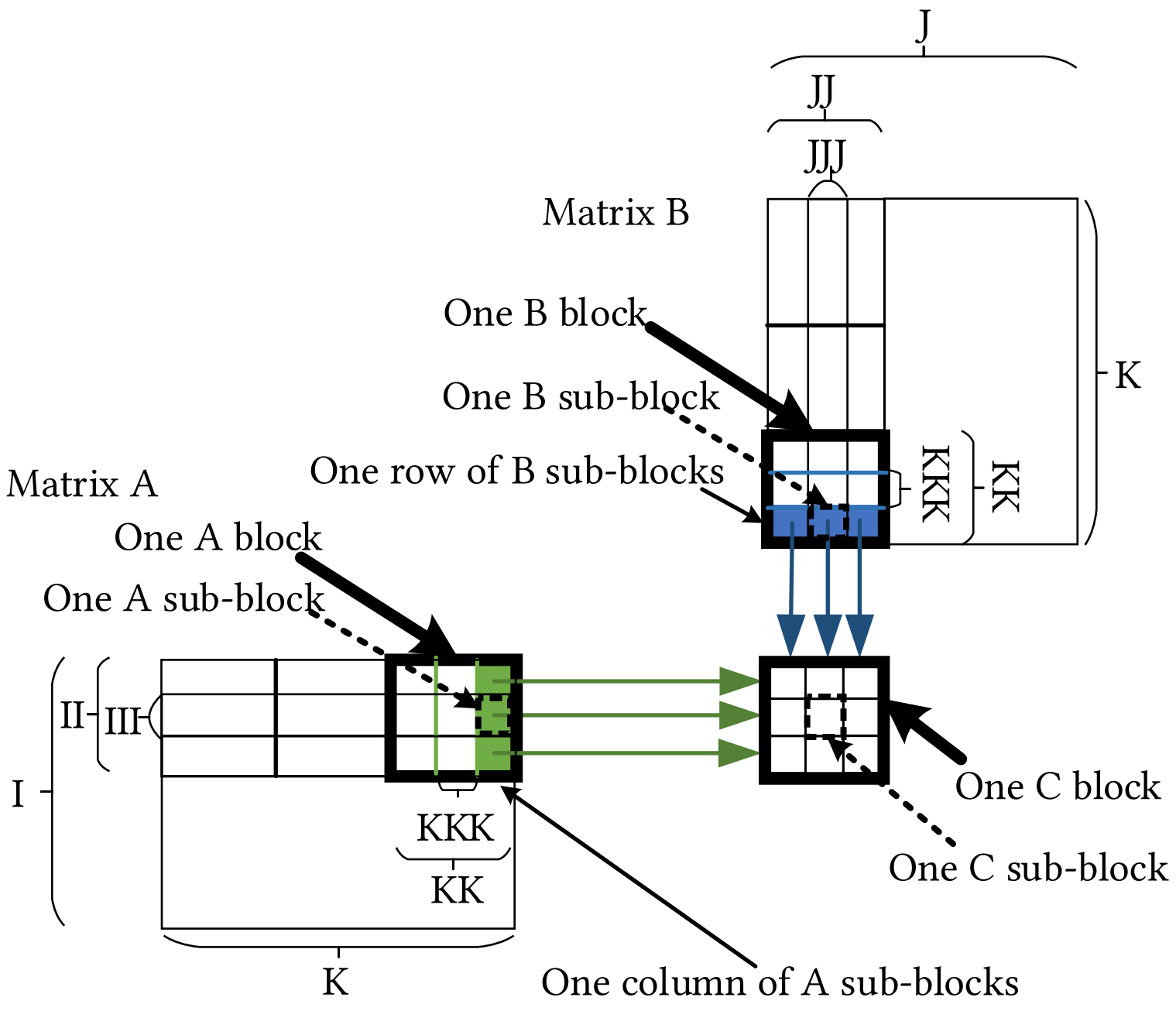

**(c) Overall idea.**

**Figure 7 SGEMM**

## Appendix

First, let us introduce some convention that clearly defines a T2S compiler's behavior.

1. *Every function, as declared as a Func, has its own loop nest.* So if there is more than one function, each function has a separate loop nest [8]. If a function is created during compute partition of another loop nest, the function's loop nest will be created by the compiler automatically.
2. *There is a channel, implicitly, between a producer and a consumer function.* If function A is called by function B, the compiler will create a channel connecting function A to function B. The channel is the output channel of function A, and is also an input channel of function B.
3. *Encapsulation of an imperative function.* Sometimes, a function has to be implemented imperatively. The imperative function is then encapsulated as a dataflow function.

   Say function F is implemented by an imperative function F_impl, we specify their correspondence using a "define_extern" directive. This directive is inherited and modified from Halide. The specification is as follows:

   *F.define_extern("F_impl", return type, parameters)*

   where return type is always void: F_impl will return results through some channel parameters, instead. The parameters are specified in his way:

   *{explicit parameters, implicit parameters}*

   The explicit parameters are instantiated explicitly in a T2S code, in the order the compiler sees them. The implicit parameters are channels, in the order the compiler creates them.

   For example, for the following T2S code,

```
Parameter A(int, 1);                  // A 1-D integer matrix
Func F1, F2, F3;                      // 3 functions
Var i;
F1(i) = F2(i, A);                     // define F1
F2.isolate_producer_chain(A, F3);  //Isolate F3 out of F2
```

   Here only function F1 is explicitly defined. Assume function F2 has an imperative implementation. Let us assume variable i is integer type and F2 returns floating-point values. We can specify F2's imperative implementation in this way:

```
F2.define_extern("F2_impl", void,
                 {int i,
                  channel int    channel1,
                  channel float channel2});
```

   The first parameter (int i) is explicit, as the compiler sees from "F2(i, A)". The two channels are implicit parameters: the first channel (channel1) is created when the compiler sees "F2.isolate_producer_chain(A, F3)"; the second channel (channel2) is created as the output channel of F2 at the end (For convenience, we assume the output channel of a function is always created the last).

   An implementation of F2 should follow the above functional interface, and be realized in an imperative language. The compiler will automatically replace any invocation of function F2 with that implementation.

   In the above example, F3 is the result of isolation. Usually, if a function F is isolated from another function that has a loop nest specified, the compiler can automatically determine the loop nest for F. But in the above example, since the original function from which F3 is isolated, i.e. function F2, is defined imperatively and has no loop nest specified, the compiler cannot determine F3's loop nest. Therefore, F3 needs to be defined imperatively as well. It can be specified as follows:

```
F3.define_extern("F3_impl", void,
                 {int A[],
                  channel int out});
```

   Here the first parameter is the integer matrix A, which appears explicitly in "F2.isolate_producer_chain(A, F3)". The compiler will automatically associate matrix A with the first parameter for the implementation to be invoked. The other parameter is implicit. It is the output channel the compiler created for F3.

Below we look at a few workloads in detail.

### A.1. Convolution and ReLU

Figure 8(a) shows the T2S code for convolution and ReLU in VGG-16, based on the design in the literature [14].

Line 1~9 of Figure 8(a) specifies the temporal definitions of the two functions. Their sequential loop nests are shown in Figure 8(b). Then a spatial mapping is specified:

1. Transform the loops.
   Two levels of tiling are performed (Line 10~13 of Figure 8(a)). The two loop nests are tiled in exactly the same way. The resulting loop nests are shown in Figure 8(c).
2. Build a basic spatial layout.
   The two loop nests are further isolated into spatial pieces (Line 14~16 of Figure 8(a)).
3. Specialize individual computations.
   ReLU is unrolled into a 2-D systolic array, and the results of ReLU are buffered at an outer loop level y (Line 17 of Figure 8(a)). Convolution is unrolled into a 3-D systolic array (Line 18). As the convolution and ReLU function have producer-consumer relationship, there is implicitly a channel between them according to our convention. As unrolling goes, the channel is split into many channels, and the PEs in the two systolic arrays are naturally connected with the channels, according to their producer-consumer relationship.
   For the input loader, some unrelated loops are removed to avoid redundant loads (Line 19). For compensation, a buffer is built at loop level y on the side of the consumer (i.e. the input feeder) (Line 20). *The loop level* (y) *at which the buffer is built immediately encloses the outermost removed loop* (noo), and thus the buffer will cache the data that are shared by all the iterations of the noo loop in the consumer side. *This illustrates a general way to remove a loop in a producer and compensate for the removal in a consumer.*

[8] The only exception is when a function is explicitly called by another function with a "compute_at" Halide directive. In this case, the function will be inlined into the caller function. This case does not appear in any of our examples, though.

Then a line buffer is further built for efficiency at another loop (Line 20 of Figure 8(a)). The buffer and the line buffer are located at an outer and an inner loop, respectively, and thus the compiler knows to fill the line buffer from the buffer. This in effect builds a two-level user-managed data cache hierarchy. *This is a general way to construct a mult-level user-managed data cache hierarchy.* Of course, in general, each level can be either a buffer or line buffer.

The loader and feeder for the weight are similar (Line 21~22 of Figure 8(a)).

Figure 8(d) shows the corresponding spatial hardware.

## A.2. SpMV

This SpMV design is based on literature [15]. The temporal definition of SpMV is shown in Line 1~7 of Figure 9(a). In this definition, we pre-split the multiplication operation out as a "product" function, because so far, we isolate functionality based on data, not on operation. So there are two functions, product and y, defined. Their corresponding loop nests are shown in Figure 9(b).

The spatial mapping is as follows:

1. Build a basic spatial layout.
   A "matrix_fetcher" is isolated out to fetch all the matrices, including A', column_ids, and row_lengths (Line 8~11 of of Figure 9(a)).
2. Specialize individual computations.
   In Line 14 of of Figure 9(a), first, the inner loop of function y is tiled with a factor NUM_ACCS, which is the number of accumulators in the design. Then the tiled inner loop is unrolled to create NUM_ACCS number of accumulators (a 1-D systolic array), each will be responsible for accumulating results for NUM_SLOTS/NUM_ACCS number of rows. Such an accumulator was named as a "fused accumulator" in the literature [15]. The accumulated results of y are buffered outside the loop nest of function y.
   In Line 15 of of Figure 9(a), the product function's inner loop is fully unrolled into another 1-D systolic array, and the input vector x are buffered outside of the loop nest of the product function.

The spatial hardware corresponding to the T2S code is shown in Figure 9(c).

As we said in Section 6.3.2, decoder is not purely functional and matrix_fetcher is not strictly a dataflow actor. So we write both of them imperatively, and encapsulate them as dataflow functions. The correspondence of the two functions to their imperative implementations is shown in Line 16~17 of Figure 9(a) using the "define_extern" directive.

Figure 10 conceptually shows the imperative implementations of the decoder and matrix_fetcher, where NUM_ROWS is the number of rows in the original matrix A.

## A.3. Merge sort

The temporal definition is shown in Line 1~7 of Figure 11(a).

In this temporal definition, the original input is a 1-D array of integer arrays (Line 1). One can vision each of the input arrays as a stream of data.

There are two symbolic constants, L and S (Line 2). L is the total number of levels in the merge sort tree. S is the size of one input array to a node at the first level. For the next level, the size of an input doubles.

Each node in the merge sort tree is located by two parameters, one for the level in the tree, the other for its position at the level. The node's level is 1 less than the level of its parent node, if any. The node's position at its level divided by 2 is its parent node's position in the next level.

Every node merges two streams of data. The merge function used in Line 5~6 has 3 parameters: the first parameter is the size of one input to the node, and the other two parameters are the input channels. This merge function will be implemented imperatively.

The original loop nest corresponding to the temporal definition is shown in Figure 11(b).

Now we define a spatial mapping:

1. Build a basic spatial layout.
   In Line 8~9 of Figure 11(a), a loader for the input is isolated out.
2. Specialize individual computations.
   In Line 10 of Figure 11(a), the output function's loop nest is fully unrolled at both loop levels. Then a PE (i.e. a tree node) forwards data toward its parent PE (i.e. the parent node of the tree node, if any). The PE is indexed by the two unrolled loop variables l and m. Its parent PE, if any, is then indexed by l+1 and floor(m/2). Therefore, the direction vector for forwarding the data is <1, floor(m/2) – m>.
   As we said, each PE merges two data streams. This merging is not functional: it compares two data items from the two streams, writes the smaller one to the output, and keeps the bigger one for the next comparison. That is, the merging has an internal state.
   So in Line 11, the merge function is specified via a define_extern directive.

The spatial hardware corresponding to the T2S code is shown in Figure 12(a). A conceptual implementation of the merge function is illustrated in Figure 12(b).

## Acknowledgement

I appreciate the valuable advice and solid support from Paul Petersen, Tim Mattson, Geoff Lowney, Chris Hughes, Pradeep Dubey, and Jim Held. The idea was developed during my working on a project with Nithin Gorge, Mike Voss, Pablo Reble, Vishakha Agrawal, Vasanth Tovinkere, and Carmen Badea for a heterogeneous programming environment. I have learned a lot when studying the expert implementations of the Programmable Solutions Group (PSG) at Intel, thanks to the help of Gordon Chiu, Tomasz Czajkowski, Davor Capalija, Dmitry Denisenko, Andrei Hagiescu, Mohamed Abdelfattah, Andrew Ling, John Freeman, and James Brodman. Many other people from various groups at Intel, including PSG, SSG, ICG and Intel Labs, have offered their much-appreciated help in various forms.

```
1  Parameter   input(int, 3), weight(int, 4), bias(int, 1), Nof, Noy, Nox, Nif, Nky, Nkx;
2  Assumption symbolic_constants(S);
3  Func          convolution, ReLU;
4  Var           no, y, x, ni, ky, kx;
5  convolution(no, y, x)   = 0;
6  convolution(no, y, x) += input(ni, y * S + ky, x * S + kx) * weight(no, ni, ky, kx);
7  ReLU(no, y, x) = max(convolution(no, y, x) +bias(no), 0);
8  convolution.set_bounds(no, 0 .. Nof).set_bounds(y, 0 .. Noy).set_bounds(x, 0 .. Nox)
              .set_bounds(ni, 0 .. Nif).set_bounds(ky, 0 .. Nky).set_bounds(kx, 0 .. Nkx);
9  ReLU         .set_bounds(no, 0 .. Nof).set_bounds(y, 0 .. Noy).set_bounds(x, 0 .. Nox);
```

Temporal definition

```
10 Assumption symbolic_constants(Tof, Toy, Pof, Poy, Pox),
              divisible(Nof, Tof), divisible(Tof, Pof),
              divisible(Noy, Toy), divisible(Toy, Poy), divisible(Nox, Pox);
11 Var           noo, yy, xx, nooo, yyy;
12 covolution.tile(no, y, noo, yy, nooo, yyy, Tof, Toy, Pof, Poy).tile(x, xx, Pox)
            .reorder(xx, nooo, yyy, yy, x, noo, y, no);
13 ReLU       .tile(no, y, noo, yy, nooo, yyy, Tof, Toy, Pof, Poy).tile(x, xx, Pox)
            .reorder(xx, nooo, yyy, yy, x, noo, y, no);
```

Transform the loops

```
14 Func          input_loader, input_feeder,
              weight_loader, weight_feeder, bias_feeder;
15 ReLU.isolate_producer_chain(bias, bias_feeder);
16 convolution.isolate_producer_chain(input, input_loader, input_feeder)
             .isolate_producer_chain(weight, weight_loader, weight_feeder);
```

Build a basic spatial layout

```
17 ReLU.unroll(xx, nooo).relay(bias, <1, 0>).buffer(ReLU, y);
18 convolution.unroll(xx, nooo, yyy).relay(input, <0, 1,0>).relay(weight, <1, 0, 0>);
19 input_loader.remove(noo, nooo);
20 input_feeder.buffer(input, y).linebuffer(input, yy);
21 weight_loader.remove(y, x, yy, xx, yyy);
22 weight_feeder.buffer(weight, no);
```

Specialize individual computations

Spatial mapping

**(a) T2S code**

```
for no in 0 .. Nof
  for y in 0 .. Noy
    for x in 0 .. Nox
      convolution(no, y, x) = 0
      for ni in 0 .. Nif
        for ky in 0 .. Nky
          for kx in 0 .. Nkx
             convolution(no, y, x) +=
               input(ni, y * S + ky, x * S + kx) * weight(no, ni, ky, kx)
```

```
for no, y, x as above
      ReLU(no, y, x) = max(convolution(no, y, x) + bias(no), 0)
```

**(b) The sequential loop nests (Line 1~9 in figure (a) above)**

```
for no in 0 .. Nof step Tof                  } First level tiling
 for y in 0 .. Noy step Toy
  for noo in 0 .. Tof step Pof               } Second level tiling
   for x in 0 .. Nox  step Pox
    for yy in 0 .. Toy step Poy
     for yyy in 0 .. Poy
      for nooo in 0 .. Pof
       for xx in 0 .. Pox
        no', y', x' = no + noo + nooo, y + yy + yyy, x + xx
        convolution(no', y', x') = 0
        for ni in 0 .. Nif
         for kx in 0 .. Nkx
          for ky in 0 .. Nky
           convolution(no', y', x') +=
             input(ni, y' * S + ky, x' * S + kx) * weight(no', ni, ky, kx)
```

```
for no, y, noo, x, yy, yyy, nooo, xx as above
        no', y', x' = no + noo + nooo, y + yy + yyy, x + xx
        ReLU(no', y', x' ) = max(convolution(no', y', x' ) + bias(no'), 0)
```

**(c) Transformed loop nests (Line 10~13 in figure (a) above)**

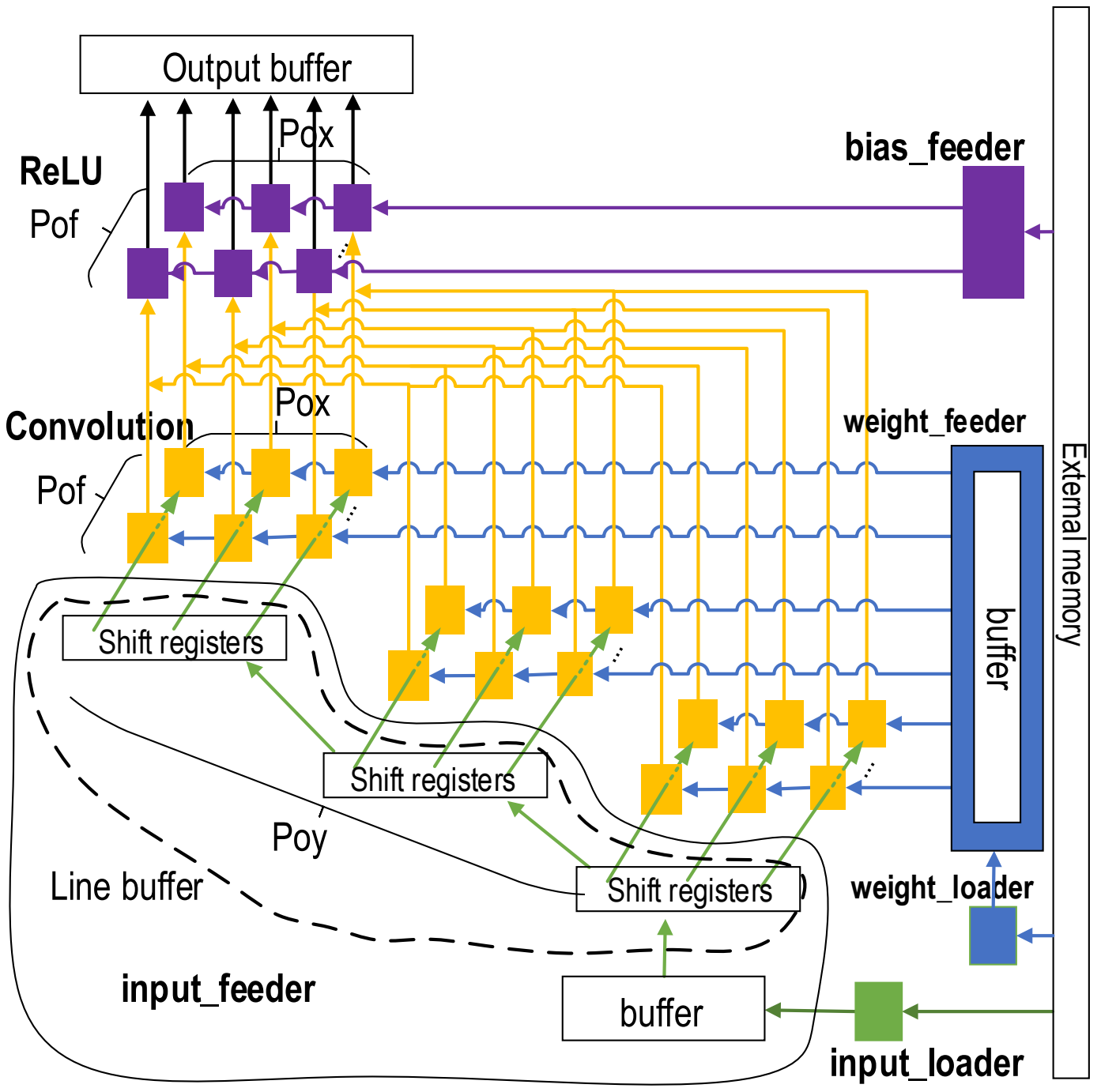

**(d) Spatial hardware**

**Figure 8 Convolution and ReLU**

```
1  Parameter   A'(float, 2), column_ids(int, 2), row_lengths(int, 2), x(float, 1), C,  NUM_ROWS;
2  Assumption symbolic_constants(NUM_SLOTS);
3  Var           c, r;
4  Func          product, decoder, y;
5  product(c, r) = A'(c, r) * x(column_ids(c, r));
6  y(decoder(row_lengths)) += product(c, r);
7  y.set_bounds(c, 0 .. C).set_bounds(r, 0 .. NUM_SLOTS);

8  Func          matrix_fetcher;
9  product.isolate_producer_chain(A', matrix_fetcher);
10 product.isolate_producer_chain(column_ids, matrix_fetcher);
11 decoder.isolate_producer_chain(row_lengths, matrix_fetcher);

12 Assumption symbolic_constants(NUM_ACCS),
              divisible(NUM_SLOTS, NUM_ACCS);
13 Var            rr;
14 y.tile(r, rr, NUM_ACCS).unroll(rr).buffer(y);
15 product.unroll(r).buffer(x);
16 decoder.define_extern( "decoder_impl", void,
                          { channel int in_row_lengths[],
                            channel int out_row_ids[]      });
17 matrix_fetcher.define_extern(  "matrix_fetcher_impl", void,
                                 {
                                               float in_A'[],
                                               int   in_column_ids[],
                                               int   in_row_lengths[],
                                    channel float out_A'[],
                                    channel int   out_column_ids[],
                                    channel int   out_row_lengths[]  });
```

Temporal definition

Build a basic spatial layout

Specialize individual computations

Spatial mapping

**(a) T2S code**

```
for c = 0 .. C
  for r = 0 .. NUM_SLOTS
    product(c, r) = A'(c, r) * x(column_ids(c, r));
```

```
for c = 0 .. C
  for r = 0 .. NUM_SLOTS
    y(decoder(row_lengths)) += product(c, r);
```

**(b) Sequential loop nests (corresponding to Line 1~7 in figure (a) above)**

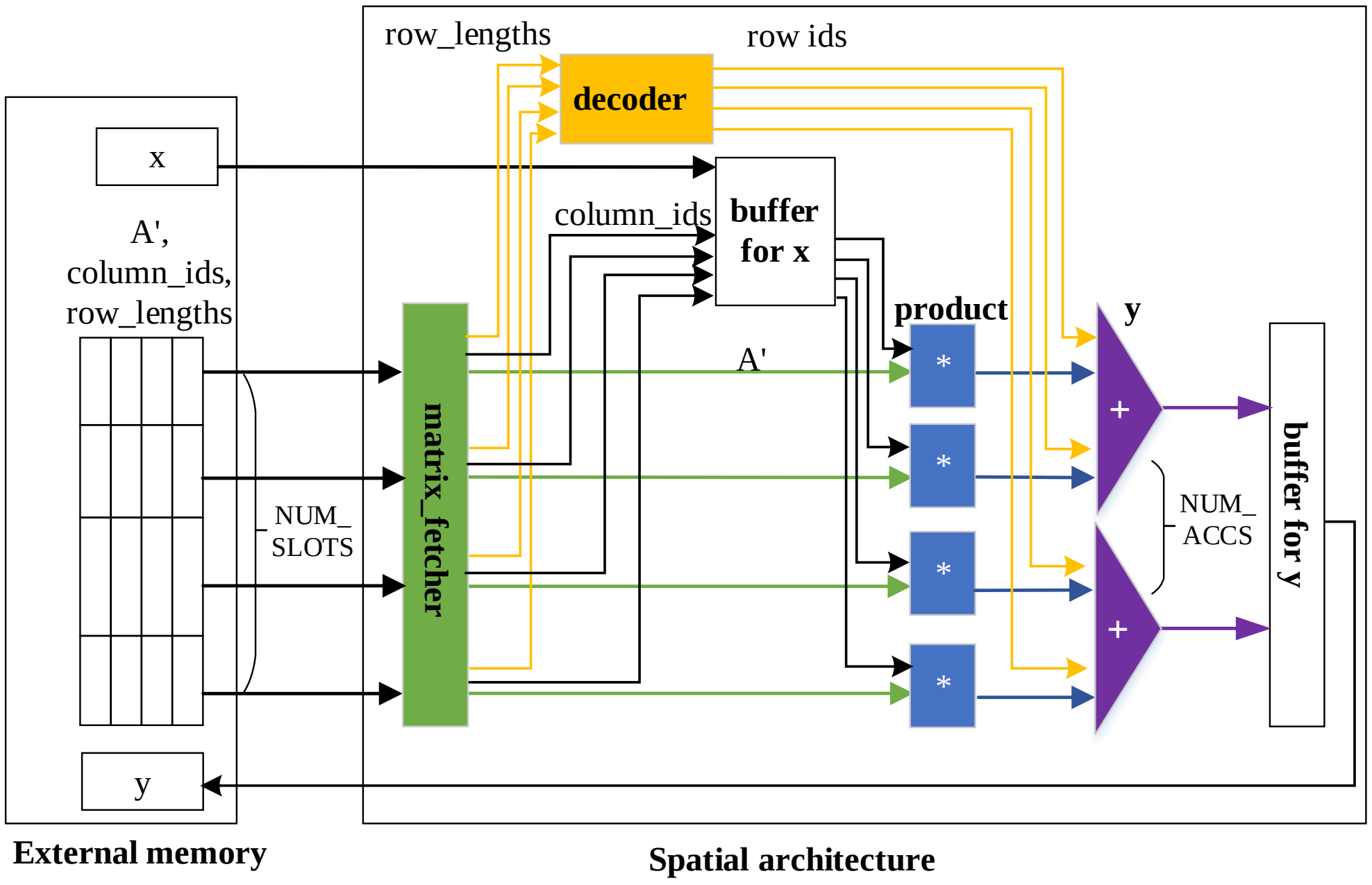

**(c) Spatial hardware**

**Figure 9. SpMV**

```
void decoder_impl(channel int in_row_lengths[],
                              channel int out_row_ids[])
{
  int counters[0 .. NUM_SLOTS] = 0;
  int row_ids[0 .. NUM_SLOTS] = 0;
  int cur_row_id = 0;
  while (1)
    for r = 0 .. NUM_SLOTS
      if (counters[r] == 0)
        read from channel in_row_lengths[r]  and write into counters[r];
        row_ids[r] = cur_row_id++;
      counters[r]--;
      write row_ids[r] into channel out_row_ids[r];
}

void matrix_fetcher_impl(           float   in_A'[],
                                     int     in_column_ids[],
                                     int     in_row_lengths[],
                            channel float  out_A'[],
                            channel int    out_column_ids[],
                            channel int    out_row_lengths[])
{
  for c = 0 .. C
    for r = 0 .. NUM_SLOTS
      write in_A'[c][r] into channel out_A'[r];
      write in_column_ids[c][r] into channel out_column_ids[r];
      if (c * NUM_SLOTS < NUM_ROWS)
        write in_row_lengths[c][r] into channel out_row_lengths[r];
}
```

**Figure 10 Imperative definitions of decoder and matrix_fetcher for SpMV in Figure 9**

```
1  Parameter    input(int*, 1);
2  Assumption symbolic_constants(L, S);
3  Func           output, merge;
4  Var             l, m;
5  output(0, m) = merge(S, input(2 * m), input(2 * m + 1));
6  output(l, m)  = merge(2^l * S, output(l - 1, 2 * m), output(l - 1, 2 * m + 1));
7  output.set_bounds(l, 0 .. L).set_bounds(m, 0 .. 2^(L - 1 - l));
                                                         Temporal definition

8  Func           loader;
9  output.isolate_producer_chain(input, loader);
                                                         Build a basic spatial layout

10 output.unroll(l, m)
            .relay(output, <1, floor(m / 2) - m>)
            .set_min_depth(output, output, 2^(l + 1) * S);
11 merge.define_extern(    "merge_impl", void,
                          {             int input_size,
                            channel int input1,
                            channel int input2,
                            channel int output
                          });
                                                         Specialize individual computations
                                                         Spatial mapping
```

**(a) T2S code**

```
for l in 0 .. L
  for m = 0 .. 2^(L-1-l)
    if (l == 0)
      out(l, m) = merge(S, input(2 * m), input(2 * m + 1))
    else
      out(l, m) = merge(2^l * S, out(l - 1, 2 * m), out(l - 1, 2 * m + 1))
```

**(b) Sequential loop nest (corresponding to Line 1~7 of figure(a) above)**

**Figure 11. Merge sort**

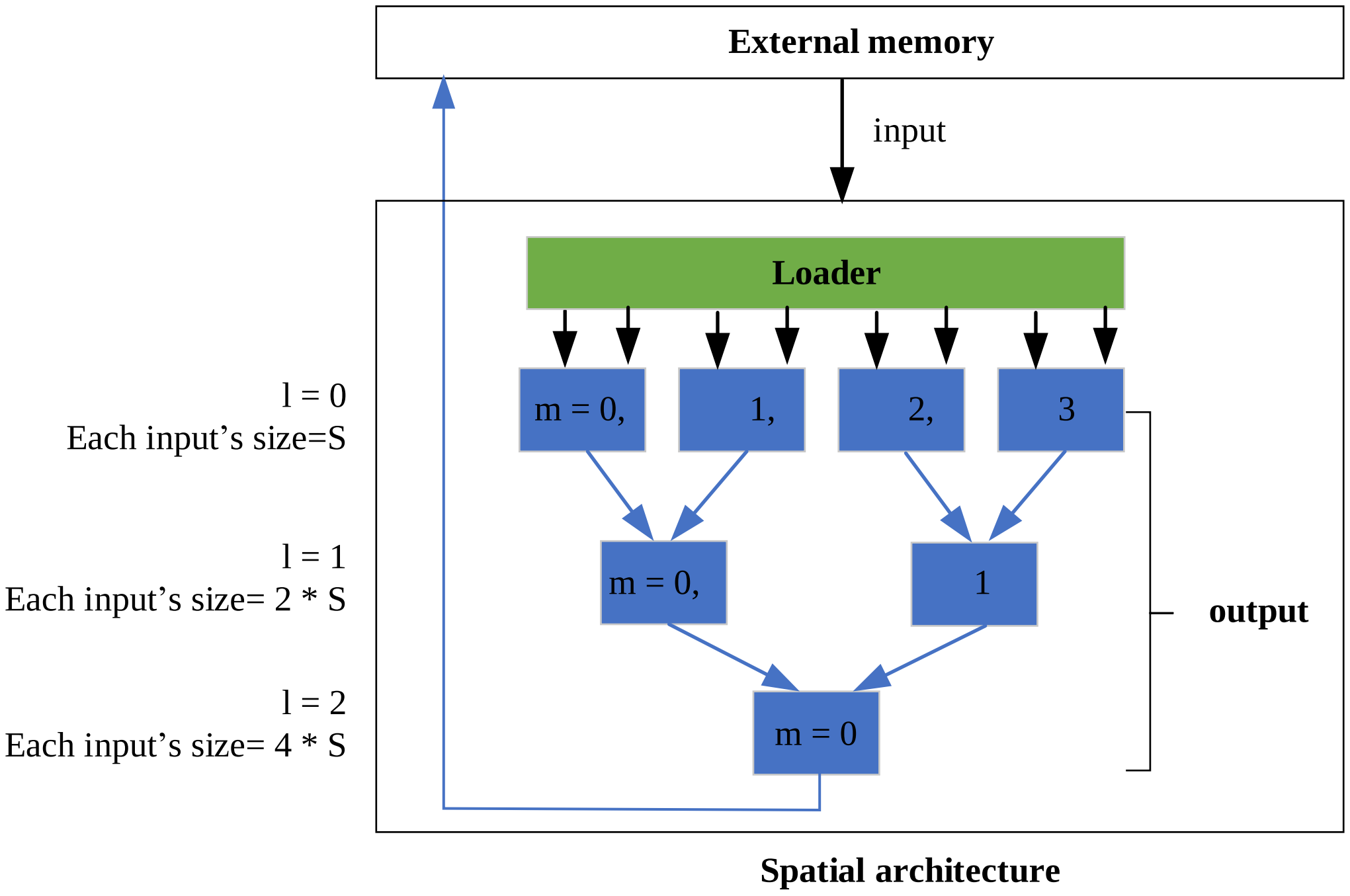

**(a) Spatial hardware for merge sort in Figure 11**

```
1  void merge_impl(              int input_size,
                            channel int input1,
                            channel int input2,
                            channel int output)
2  {
3      int a = read from the input1 channel ;
4      int b = read from the input2 channel;
5      int idx1 = idx2 = 1;
6      for (int i = 0; i < 2 * input_size; i++) {
7         if( idx1 ≥ input_size || (idx2 < input_size && a ≥ b))
8             write b into the output channel; b = read from the input2 channel; idx2++;
9         else
10            write a into the output channel; a = read from the input1 channel; idx1++;
11  }
```

**(b) Imperative definition of merge in Figure 11**

**Figure 12 Spatial hardware and imperative definition for merge sort**

## References

1 Altera white paper. Enabling high-performance floating-point designs. https://www.altera.com/content/dam/altera-www/global/en_US/pdfs/literature/wp/wp-01267-fpgas-enable-high-performance-floating-point.pdf

2 Wikipedia. Field-programmable gate array. https://en.wikipedia.org/wiki/Field-programmable_gate_array

3 Tony Nowatzki, Vinay Gangadhar, Newsha Ardalani, and Karthikeyan Sankaralingam. 2017. Stream-Dataflow Acceleration. *SIGARCH Comput. Archit. News* 45, 2 (June 2017), 416-429.

4 Mihai Budiu, Girish Venkataramani, Tiberiu Chelcea, and Seth Copen Goldstein. 2004. Spatial computation. *SIGPLAN Not.* 39, 11 (October 2004), 14-26.

5 Reiner Hartenstein. 2001. Coarse grain reconfigurable architecture (embedded tutorial). In *Proceedings of the 2001 Asia and South Pacific Design Automation Conference (ASP-DAC '01).* ACM, New York, NY, USA, 564-570.

6 Vaishali Tehre and Ravindra Kshirsagar. Survey on Coarse Grained Reconfigurable Architectures. *International Journal of Computer Applications* 48(16):1-7, June 2012.

7 Angshuman Parashar, Michael Pellauer, Michael Adler, Bushra Ahsan, Neal Crago, Daniel Lustig, Vladimir Pavlov, Antonia Zhai, Mohit Gambhir, Aamer Jaleel, Randy Allmon, Rachid Rayess, Stephen Maresh, and Joel Emer. 2013. Triggered instructions: a control paradigm for spatially-programmed architectures. In *Proceedings of the 40th Annual International Symposium on Computer Architecture* (ISCA '13). ACM, New York, NY, USA, 142-153.

8 High level synthesis. https://en.wikipedia.org/wiki/High-level_synthesis

9 David Bacon, Rodric Rabbah, and Sunil Shukla. 2013. FPGA Programming for the Masses. *Queue* 11, 2, pages 40 (February 2013), 13 pages.

10 Nadathur Satish, Changkyu Kim, Jatin Chhugani, Hideki Saito, Rakesh Krishnaiyer, Mikhail Smelyanskiy, Milind Girkar, and Pradeep Dubey. 2012. Can traditional programming bridge the Ninja performance gap for parallel computing applications?. In *Proceedings of the 39th Annual International Symposium on Computer Architecture (ISCA '12).* IEEE Computer Society, Washington, DC, USA, 440-451.

11 Jonathan Ragan-Kelley, Connelly Barnes, Andrew Adams, Sylvain Paris, Frédo Durand, and Saman Amarasinghe. 2013. Halide: a language and compiler for optimizing parallelism, locality, and recomputation in image processing pipelines. In *Proceedings of the 34th ACM SIGPLAN Conference on Programming Language Design and Implementation (PLDI '13).* ACM, New York, NY, USA, 519-530.

12 Jing Pu, Steven Bell, Xuan Yang, Jeff Setter, Stephen Richardson, Jonathan Ragan-Kelley and Mark Horowitz. Programming Heterogeneous Systems from an Image Processing DSL. CoRR, abs/1610.09405, 2016. http://arxiv.org/abs/1610.09405.

13 Altera white paper. Accelerating Genomics Research with OpenCL and FPGAs. https://www.altera.com/content/dam/altera-www/global/en_US/pdfs/literature/wp/wp-01262-accelerating-genomics-research-with-opencl-and-fpgas.pdf

14 Yufei Ma, Yu Cao, Sarma Vrudhula, and Jae-sun Seo. 2017. Optimizing Loop Operation and Dataflow in FPGA Acceleration of Deep Convolutional Neural Networks. In *Proceedings of the 2017 ACM/SIGDA International Symposium on Field-Programmable Gate Arrays (FPGA '17).* ACM, New York, NY, USA, 45-54.

15 J. Fowers, K. Ovtcharov, K. Strauss, E. S. Chung and G. Stitt. A High Memory Bandwidth FPGA Accelerator for Sparse Matrix-Vector Multiplication. In *Proceedings of 2014 IEEE 22nd Annual International Symposium on Field-Programmable Custom Computing Machines*, Boston, MA, 2014, pp. 36-43.

16 Dirk Koch and Jim Torresen. 2011. FPGASort: a high performance sorting architecture exploiting run-time reconfiguration on fpgas for large problem sorting. In *Proceedings of the 19th ACM/SIGDA international symposium on Field programmable gate arrays (FPGA '11).* ACM, New York, NY, USA, 45-54.

17 Samuel Williams, Andrew Waterman, and David Patterson. 2009. Roofline: an insightful visual performance model for multicore architectures. *Commun. ACM* 52, 4 (April 2009), 65-76.

18 Jason Cong, Peng Wei, Cody Hao Yu, and Peipei Zhou. Bandwidth Optimization Through On-Chip Buffer Restructuring for HLS . Proceedings of the 54rd Annual Design Automation Conference (DAC 2017), Austin, TX, June 18-22, 2017.

19 Xuechao Wei, Cody Hao Yu, Peng Zhang, Youxiang Chen, Yuxin Wang, Han Hu, Yun Liang, and Jason Cong. Automated Systolic Array Architecture Synthesis for High Throughput CNN Inference on FPGAs . Proceedings of the 54rd Annual Design Automation Conference (DAC 2017), Austin, TX, June 18-22, 2017.

20 Chen Zhang, Peng Li, Guangyu Sun, Yijin Guan, Bingjun Xiao, and Jason Cong. 2015. Optimizing FPGA-based Accelerator Design for Deep Convolutional Neural Networks. In *Proceedings of the 2015 ACM/SIGDA International Symposium on Field-Programmable Gate Arrays* (FPGA '15). ACM, New York, NY, USA, 161-170.

21 J. J. Navarro, J. M. Llaberia and M. Valero, "Partitioning: An Essential Step in Mapping Algorithms Into Systolic Array Processors," in *Computer*, vol. 20, no. 7, pp. 77-89, July 1987.

22 Krste Asanović, Ras Bodik, Bryan Christopher Catanzaro, Joseph James Gebis, Parry Husbands, Kurt Keutzer, David A. Patterson, William Lester Plishker, John Shalf, Samuel Webb Williams and Katherine A. Yelick. The Landscape of Parallel Computing Research: A View from Berkeley. EECS Department, University of California, Berkeley. *Technical Report* No. UCB/EECS-2006-183. December 18, 2006

23 Kurt Keutzer, Berna L. Massingill, Timothy G. Mattson, and Beverly A. Sanders. 2010. A design pattern language for engineering (parallel) software: merging the PLPP and OPL projects. In *Proceedings of the 2010 Workshop on Parallel Programming Patterns (ParaPLoP '10)*. ACM, New York, NY, USA, Article 9, 8 pages.

24 C. Karfa, D. Sarkar and C. Mandal, "Verification of KPN Level Transformations," *2013 26th International Conference on VLSI Design and 2013 12th International Conference on Embedded Systems*, Pune, 2013, pp. 338-343.

25 Frank Hannig, Hritam Dutta, and Jurgen Teich, Mapping of regular nested loop programs to coarse-grained reconfigurable arrays - constraints and methodology, In *Proceedings of the 18th International Parallel and Distributed Processing Symposium, 2004.* Santa Fe, NM, USA, 2004, pp. 148-.

26 Utpal K. Banerjee. 1993. Loop Transformations for Restructuring Compilers: The Foundations. Kluwer Academic Publishers, Norwell, MA, USA.

27 H. T. Kung. 1982. Why Systolic Architectures? Computer 15, 1 (January 1982), 37-46.

28 Sun-Yuan Kung, K. S. Arun, R. J. Gal-Ezer, and D. V. Bhaskar Rao. 1982. Wavefront Array Processor: Language, Architecture, and Applications. IEEE Trans. Comput. 31, 11 (November 1982), 1054-1066.

29 Hervé Le Verge, Christophe Mauras, Patrice Quinton. The ALPHA language and its use for the design of systolic arrays. Journal of VLSI signal processing systems for signal, image and video technology, September 1991, Volume 3, Issue 3, pp 173–182

30 Wang Luzhou, Kentaro Sano, and Satoru Yamamoto. 2012. Domain-Specific language and compiler for stencil computation on FPGA-Based systolic computational-memory array. In *Proceedings of the 8th international conference on Reconfigurable Computing: architectures, tools and applications (ARC'12)*, Oliver S. Choy, Ray C. Cheung, Peter Athanas, and Kentaro Sano (Eds.). Springer-Verlag, Berlin, Heidelberg, 26-39.

31 Yuan Tang, Rezaul Alam Chowdhury, Bradley C. Kuszmaul, Chi-Keung Luk, and Charles E. Leiserson. 2011. The pochoir stencil compiler. In *Proceedings of the twenty-third annual ACM symposium on Parallelism in algorithms and architectures (SPAA '11).* ACM, New York, NY, USA, 117-128.

32 Ravi Teja Mullapudi, Vinay Vasista, and Uday Bondhugula. 2015. PolyMage: Automatic Optimization for Image Processing Pipelines. SIGARCH Comput. Archit. News 43, 1 (March 2015), 429-443.

33 James Hegarty, John Brunhaver, Zachary DeVito, Jonathan Ragan-Kelley, Noy Cohen, Steven Bell, Artem Vasilyev, Mark Horowitz, and Pat Hanrahan. 2014. Darkroom: compiling high-level image processing code into hardware pipelines. ACM Trans. Graph. 33, 4, Article 144 (July 2014), 11 pages.

34 Nithin George, HyoukJoong Leey, David Novo, Tiark Rompf, Kevin J. Browny, Arvind K. Sujeethy, Martin Odersky, Kunle Olukotuny and Paolo Ienne. "Hardware system synthesis from Domain-Specific Languages," 2014 24th International Conference on Field Programmable Logic and Applications (FPL), Munich, 2014, pp. 1-8

35 James Hegarty, Ross Daly, Zachary DeVito, Jonathan Ragan-Kelley, Mark Horowitz, and Pat Hanrahan. 2016. Rigel: flexible multi-rate image processing hardware. ACM Trans. Graph. 35, 4, Article 85 (July 2016), 11 pages.

36 Circular shift. https://en.wikipedia.org/wiki/Circular_shift

37 Wikipedia. Communicating Sequential Processes. https://en.wikipedia.org/wiki/Communicating_sequential_processes

38 Jack Dennis. "Data Flow Graphs." *Encyclopedia of Parallel Computing*. Ed. David Padua. Boston, MA : Springer US, 2011. 512--518.

39 Heinrich Riebler, Michael Lass, Robert Mittendorf, Thomas Löcke, and Christian Plessl. 2017. Efficient Branch and Bound on FPGAs Using Work Stealing and Instance-Specific Designs. ACM Trans. Reconfigurable Technol. Syst. 10, 3, Article 24 (June 2017), 23 pages.

40 E. Del Sozzo, L. Di Tucci and M. D. Santambrogio, "A highly scalable and efficient parallel design of N-body simulation on FPGA," In *Proceddings of 2017 IEEE International Parallel and Distributed Processing Symposium Workshops (IPDPSW)*, Lake Buena Vista, FL, 2017, pp. 241-246.

41 Kenji Kanazawa and Tsutomu Maruyama. 2010. An Approach for Solving Large SAT Problems on FPGA. ACM Trans. Reconfigurable Technol. Syst. 4, 1, Article 10 (December 2010), 21 pages.

42 Xiang Tian and K. Benkrid, "Design and implementation of a high performance financial Monte-Carlo simulation engine on an FPGA supercomputer," *2008 International Conference on Field-Programmable Technology,* Taipei, 2008, pp. 81-88.

43 Brahim Betkaoui, Yu Wang, David B. Thomas, and Wayne Luk. 2012. A Reconfigurable Computing Approach for Efficient and Scalable Parallel Graph Exploration. In *Proceedings of the 2012 IEEE 23rd International Conference on Application-Specific*

*Systems, Architectures and Processors* (ASAP '12). IEEE Computer Society, Washington, DC, USA, 8-15.
44 Kyrylo Tkachov. Accelerating Unstructured Mesh. Computations using Custom Streaming. Architectures. *Project report.* Department of Computing, Imperial College London, 2012.
45 Becker T., Mencer O., Gaydadjiev G. (2016) Spatial Programming with OpenSPL. In: Koch D., Hannig F., Ziener D. (eds) *FPGAs for Software Programmers.* Springer, Cham.
46 Intel. Intel® FPGA SDK for OpenCL Programming Guide. 2017.
47 Jan Decaluwe. 2004. MyHDL: a python-based hardware description language. Linux J. 2004, 127 (November 2004), 5-.
48 Jonathan Bachrach, Huy Vo, Brian Richards, Yunsup Lee, Andrew Waterman, Rimas Avižienis, John Wawrzynek, and Krste Asanović. 2012. Chisel: constructing hardware in a Scala embedded language. In Proceedings of the 49th Annual Design Automation Conference (DAC '12). ACM, New York, NY, USA, 1216-1225.
49 J. Cong, B. Liu, S. Neuendorffer, J. Noguera, K. Vissers and Z. Zhang. High-Level Synthesis for FPGAs: From Prototyping to Deployment. IEEE Transactions on Computer-Aided Design of Integrated Circuits and Systems, Volume 30, Number 4, pp. 473-491, April 2011. (Keynote paper).
50 Wikipedia. Xilinx Vivado. https://en.wikipedia.org/wiki/Xilinx_Vivado
51 Jason Ansel, Shoaib Kamil, Kalyan Veeramachaneni, Jonathan Ragan-Kelley, Jeffrey Bosboom, Una-May O'Reilly, and Saman Amarasinghe. 2014. OpenTuner: an extensible framework for program autotuning. In *Proceedings of the 23rd international conference on Parallel architectures and compilation* (PACT '14). ACM, New York, NY, USA, 303-316.